\documentclass[aps,prb,twocolumn,superscriptaddress,showpacs,floatfix,longbibliography]{revtex4-1}
\pdfoutput=1
\usepackage{amsmath,amssymb,amsfonts,float,graphics,epsfig,epstopdf,color,verbatim,tabularx,bm,multirow,appendix,hyperref}
\usepackage[normalem]{ulem}
\usepackage{lmodern}
\usepackage{ytableau}
\usepackage{pifont}
\usepackage{color}
\usepackage{bm}
\usepackage{wasysym}
\DeclareMathOperator{\Tr}{Tr}
\usepackage{ dsfont }
\usepackage{graphicx}

\def\k{{\mathbf{k}}}
\def\R{{\mathbf{R}}}
\def\Q{{\mathbf{Q}}}
\def\q{{\mathbf{q}}}

\def\r{{\mathbf{r}}}

\begin{document}

\title{Strange metal and Fermi arcs from disordering spin stripes}

\author{Xu Zhang}
\affiliation{Department of Physics and Astronomy, Ghent University, Krijgslaan 281, 9000 Gent, Belgium}

\author{Nick Bultinck}
\affiliation{Department of Physics and Astronomy, Ghent University, Krijgslaan 281, 9000 Gent, Belgium}

\begin{abstract}
We revisit the effective theory for fluctuating spin stripes coupled to a Fermi surface, and consider the parameter regime where a spin nematic phase intervenes between the spin density wave state and the symmetric state. It is shown that adding potential disorder to this theory, which acts as an unconventional type of random-field disorder, naturally gives rise to a phase diagram containing a quantum critical point that is described by the universal theory of strange metals with spatial disorder in both the magnitude and sign of the electron-boson coupling term [\href{https://www.science.org/doi/abs/10.1126/science.abq6011}{A.A. Patel, H. Guo, I. Esterlis and S. Sachdev, Science \textbf{381}, 790 (2023)}]. One difference compared to the original theory, however, is that at non-zero temperatures the disordered spin-stripe model automatically self-averages over the sign of the coupling. We also study the effects of thermal fluctuations in a phenomenological model for the disordered spin density wave state, and find from Monte Carlo simulations that a short anti-ferromagnetic correlation length (order 4-5 lattice constants) already leads to pronounced Fermi arcs in the electronic spectral weight.
\end{abstract}
\date{\today}
\maketitle

\section{Introduction} 
In the underdoped region and at somewhat elevated temperatures the phase diagram of the cuprate high-temperature superconductors famously hosts a pseudogap metal. The mechanism behind the pseudogap phenomenon is much debated, but even people following different schools of thought seem to agree on the fact that understanding the nature of the pseudogap is essential for developing a comprehensive picture of the key physics shaping the cuprate phase diagram. 

In experiment one frequently observes static or fluctuating stripes, i.e. uni-directional charge or spin density wave orders, in the pseudogap region~\cite{Fradkin2015}. Originally, such stripe order was first found in numerical Hartree-Fock studies~\cite{Zaanen1989,Machida1989,Kato1990}. However, one important discrepancy is that the mean-field stripes are insulating, whereas the experimentally observed stripes are metallic. From a more strong-coupling perspective, the stripe formation was explained as a form of phase separation in regions with low and high hole density~\cite{Emery1993}. The key mechanism behind this strong-coupling theory is the interplay between two different forms of frustration: first there is the frustration of hole motion in an anti-ferromagnetic background which is driving the phase separation, and secondly there is the long-range Coulomb interaction frustrating macroscopic phase separation~\cite{Emery1993}. But irrespective of the theoretical framework used to explain stripe formation, one always finds that when spin stripes occur, they intertwine in a non-trivial way with charge stripes. In particular, the charge stripes act as anti-phase domains in the spin pattern~\cite{Zaanen1989,Zachar1998,Zaanen2001}. This means that the charge troughs, corresponding to extended one-dimensional hole-rich regions, act as domain walls in the anti-ferromagnetic spin pattern: on crossing the charge trough the up and down spins change sublattice. 

Another central feature of the cuprate phase diagram is the strange metal behavior around optimal doping. One of the hallmarks of the strange metal is the linear-in-temperature resistivity down to very low temperatures when the superconducting dome is suppressed. Such behavior, which is at odds with Fermi liquid theory, has proven challenging to reproduce theoretically in a robust way. Recently, a new theory was developed to explain the strange metal phenomenology based on the much-studied model of a Fermi surface coupled to a quantum critical boson, but with spatial randomness in the electron-boson coupling~\cite{Aldape2022,Esterlis2021,Guo2022,Patel2023}. This type of spatial randomness was argued to be crucial for obtaining the linear-T resistivity.

In this work we connect the physics of spin stripes to the strange metal theory of Refs.~\cite{Aldape2022,Esterlis2021,Guo2022,Patel2023}. For this we draw inspiration from an insightful paper by Mross and Senthil about the effects of disorder on Spin Density Wave (SDW) states~\cite{Mross2015}. Similarly to these authors, we consider a situation in which the spin stiffness is considerably larger than the charge stiffness (the latter can even be zero). In this context, spatial disorder leads to a SDW glass~\cite{Mross2015}, which we argue connects naturally to the above-mentioned strange metal theory. 

An interesting feature of the SDW glass is the emergence of Ising spin degrees of freedom. In the final part of this work we study the effects of thermal fluctuations of these Ising variables. We use a simple effective model that can be simulated with Monte Carlo on large system sizes, and find that the low-energy electronic spectral weight exhibits clear Fermi arcs even though the anti-ferromagnetic correlation length is only a few lattice constants. This brings us full circle, i.e. back to the phenomenology of the pseudogap.

The main results of this work are organized in three sections. In Sec.~\ref{sec:clean} we introduce the effective model, and discuss the phase diagram we consider in the absence of disorder. The effects of disorder on this phase diagram are discussed in Sec.~\ref{sec:dis}. Finally, in Sec.~\ref{sec:arcs} we introduce the phenomenological model to study thermal fluctuations of the emergent Ising degrees of freedom and present results on the electronic spectral weight. We end with a short discussion in Sec.~\ref{sec:disc}.

\section{Effective theory for fluctuating spin stripes}\label{sec:clean}

Our analysis is based on an effective theory for fluctuating spin stripes coupled to electrons. To construct this theory we start from the order parameter for an anti-ferromagnetic spin stripe or SDW, which is given by
\begin{equation}\label{eqSDW}
(-1)^{r_x+r_y}\langle \vec{S}(\r)\rangle = \vec{M}\cos(Qr_x)\,,
\end{equation}
where $\r = (r_x, r_y)$ are integer square-lattice coordinates. Note that we write coordinate and momentum vectors in boldface, and all other vectors (such as order parameter vectors) with an arrow. In what follows we will work with the Fourier components of the order parameter:
\begin{equation}\label{eqorder}
(-1)^{r_x+r_y} \langle \vec{S}(\r)\rangle = e^{iQr_x}\vec{S}_Q + e^{-iQr_x}\vec{S}_{-Q}\,,
\end{equation}
where $\vec{S}_{-Q}=\vec{S}_{Q}^*$. Under a translation in the $x$-direction the Fourier components transform as
\begin{equation}
T_x : \;\;\vec{S}_Q  \rightarrow e^{iQ}\vec{S}_Q \,,
\end{equation}
where we have used units in which the lattice spacing is equal to one. A spin rotation acts in the obvious way:
\begin{equation}
O(3) : \;\;\vec{S}_Q  \rightarrow O\vec{S}_Q \,,
\end{equation}
where $O$ is a $3\times 3$ real orthogonal matrix. We will formulate the effective theory in terms of the $2\times 3$ real matrix field $\Phi$, whose components are defined by
\begin{equation}\label{defPhi}
S^j_Q = \Phi_{1j} + i \Phi_{2j}
\end{equation}
When $Q$ is incommensurate, $T_x$ generates an O(2) symmetry acting on the row index of $\Phi$. Spin rotations, on the other hand, act as O(3) rotations on the column index. Crucially, the order parameter manifold is $\left( S^1\times S^2\right)/\mathbb{Z}_2$~\cite{Zaanen2001,Sachdev2002,Nussinov2002,Zhang2002,Mross2012,Mross2012PRB}. This is because when acting on $\Phi$, a $\pi$-rotation of O(2) is equivalent to an inversion of the spin indices. As a result, a pair of points on $S^1$ and $S^2$ should be identified with the pair of opposite points. The quotient changes the fundamental group of the order parameter manifold from $\mathbb{Z}$ to $\mathbb{Z}/2$, as there exists a non-contractible closed path connecting opposite points on both $S^1$ and $S^2$. This path allows for a point defect corresponding to a half O(2) vortex where the spin direction rotates from $\vec{m}$ to $-\vec{m}$ when moving around the singularity \cite{Zaanen2001,Kruger2002,Mross2012,Mross2012PRB}. This half-vortex is similar to the half-vortices of pair-density wave superconductors \cite{Agterberg2008,Berg2009,Radzihovsky2009}.

A Landau-Ginzburg analysis~\cite{Zachar1998} shows that a SDW of the form in Eq. \eqref{eqSDW} induces a Charge Density Wave (CDW) with Fourier components
\begin{equation}
\rho_{2Q} \propto \vec{S}_Q \cdot \vec{S}_Q
\end{equation}
We will not explicitly take this CDW into account with an independent field, but instead let it be generated by coupling $\Phi$ to electrons.

In the absence of disorder, we write the Euclidean action of our 2+1-dimensional effective theory as a sum of two terms:
\begin{equation}
S_{\text{clean}} = S_\psi + S_\Phi
\end{equation}
The first term is given by
\begin{eqnarray}
 S_\psi & = & \int \mathrm{d}\tau \sum_{\k} \bar{\psi}(\k)(\partial_\tau + \varepsilon_\k)\psi(\k) \\
 & + & g\int \mathrm{d}\tau \sum_{\r}  \, (-1)^{r_x+r_y}l^a(\r)\Phi_{aj}(\r) \bar{\psi}(\r)\sigma^j \psi(\r)  \nonumber \,,
\end{eqnarray}
where summation over repeated indices is implicit. Here we have introduced the spinful fermions $\psi(\r) = (\psi_\uparrow(\r),\psi_\downarrow(\r))^T$ with dispersion $\varepsilon_\k$. In the coupling between $\Phi$ and the fermions we have defined the two-dimensional vector
\begin{equation}
\vec{l}(\r) = \left(\begin{matrix} \cos(Qr_x) \\ \sin(Qr_x) \end{matrix}\right)
\end{equation}
The coupling is such that a translation $(r_x,r_y)\rightarrow (r_x+n,r_y+m)$ combined with an O(2) rotation of $\Phi$ over an angle $-Qn+\pi(n+m)$ is a symmetry of the action. Note that we are implicitly considering orthorhombic systems, but the generalization to tetragonal systems is straightforward by extending $\Phi$ to a pair of $2\times 3$ real matrix fields $(\Phi^x,\Phi^y)$ such that the $C_{4v}$ symmetry is restored.

The purely bosonic part of the effective action is given by
\onecolumngrid
\begin{eqnarray}\label{SPhi}
 S_\Phi
 & = & J \int\mathrm{d}\tau \sum_\r \frac{1}{v^2}\left(\partial_\tau\Phi_{aj}\right)^2  + \sum_{\langle \r\r'\rangle} \left(\Phi_{aj}(\r) - \Phi_{aj}(\r') \right)^2 \\
 & + & \tilde{J} \int\mathrm{d}\tau \sum_\r \frac{1}{\tilde{v}^2}\left(\partial_\tau\left[\Phi^T\Phi\right]_{jl}\right)^2+ \sum_{\langle \r\r'\rangle} \bigg( [\Phi^T\Phi]_{jl}(\r) - [\Phi^T\Phi]_{jl}(\r') \bigg)^2 \nonumber \\
 & + & \int\mathrm{d}\tau\,\sum_\r  \left[- r \,\text{tr}(\Phi^T\Phi)+\frac{u}{4} \left(\text{tr}(\Phi^T\Phi)\right)^2 -\sum_{i=x,z}\frac{v}{4} \left(\text{tr}( \Phi^T \tau^i \Phi)\right)^2  \right] \nonumber\,,
\end{eqnarray}
\twocolumngrid
where again summation over repeated indices is implicit, and $\tau^i$ are the Pauli matrices acting on the two-dimensional row index of $\Phi$. The field potential in Eq.~\eqref{SPhi} is the same as in Ref.~\cite{Sachdev2002}. We take $r,v>0$ (and $u>v$ for stability), such that this potential is minimized by
\begin{equation}
\Phi_{aj} = |\Phi| n_a m_j\,,
\end{equation}
with $n_a$ and $m_j$ the components of respectively a two- and three-dimensional unit vector. Via Eq.~\eqref{defPhi}, this rank-one $\Phi$ corresponds to a $\vec{S}_Q$ with parallel real and imaginary parts, which indeed produces the real-space SDW order in Eq. \eqref{eqSDW}. We will consider the situation where $r,u$ and $v$ are much larger than $J$ and $\tilde{J}$, such that we can consider a nonlinear sigma model for $n_am_j$ (as $|\Phi|$ can be absorbed in $J$, $\tilde{J}$ and $g$, we will without loss of generality take $|\Phi|=1$). Because the order parameter manifold is $(S^1\times S^2)/\mathbb{Z}_2$, the effective theory for $\vec{n}$ and $\vec{m}$ has a $\mathbb{Z}_2$ gauge redundancy corresponding to simultaneously flipping the sign of both $\vec{n}$ and $\vec{m}$~\cite{Zaanen2001,Sachdev2002,Nussinov2002,Zhang2002,Mross2012,Mross2012PRB}. 

The first line in Eq.~\eqref{SPhi} is a conventional kinetic or stiffness term for the order parameter field $\Phi$. The second line is a stiffness term for the (gauge-invariant) composite field
\begin{equation}
\left[\Phi^T \Phi \right]_{jl} = m_j m_l
\end{equation}
The term in the second line of Eq.~\eqref{SPhi} thus raises the action for order parameter configurations where the spin field $\vec{m}$ varies in space and time, but is insensitive to variations of the CDW field $\vec{n}$. When $\tilde{J}$, the coefficient of this term, becomes large this will lead to order for the $m_jm_l$ field, such that
\begin{equation}
Q_{jl} = \langle m_j m_l\rangle - \frac{1}{3} \delta_{jl}
\end{equation}
is non-zero. As $Q_{jl}$ transforms non-trivially under spin rotations but is invariant under $\vec{m}\rightarrow -\vec{m}$, this is an order parameter for spin nematicity.

\begin{figure}
  \includegraphics[scale=0.65]{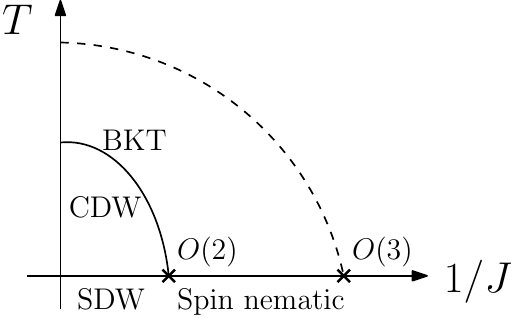}
  \caption{Phase diagram of the effective spin stripe model in the absence of disorder. The two crosses at zero temperature indicate continuous quantum phase transitions in the metallic O(2) and conventional O(3) universality classes. The solid line at non-zero temperature is a Berezinskii-Kosterlitz-Thouless (BKT) transition for the incommensurate CDW order. The dashed line represents a cross-over region where the spin nematic correlation length becomes significantly larger than the lattice constant.}
  \label{CleanPD}
\end{figure}

In this work we are interested in the following path in parameter space. We start with both $J$ and $\tilde{J}$ sufficiently large, resulting in SDW order at zero temperature, i.e. $\langle \Phi\rangle \neq 0$. We then decrease $J$ (but keep $\tilde{J}$ constant), such that fluctuations in $\vec{n}$ destroy the SDW order and as a result $\langle \Phi\rangle = 0$. We assume that $\tilde{J}$ is sufficiently large such that there is a range of $J$ for which there is no SDW order, but $Q_{jl}$ nevertheless remains non-zero. In this parameter range, the system is in a spin nematic state. Finally, for sufficiently small $J$, also $Q_{jl}=0$ and the system is fully disordered. It is well-known that fluctuating SDW order can induce spin nematic order \cite{Zaanen2001,Kruger2002,Mross2012,Mross2012PRB,Sachdev2002,Nussinov2002,Zhang2002}, and in the appendix we confirm via quantum Monte Carlo simulations that for a sign problem-free version of the effective theory considered here there indeed exists a value for $\tilde{J}$ such that decreasing $J$ leads to the phase diagram discussed above. 

The full phase diagram we consider is shown in Fig.~\ref{CleanPD}. At non-zero temperature, spin rotation symmetry is restored, and the SDW order becomes a quasi-long range CDW (assuming incommensurate $Q$). This algebraic CDW state is separated from the disordered phase by a Berezinskii-Kosterlitz-Thouless (BKT) transition. The spin nematic state is disordered at non-zero temperatures. The nematic correlation length becomes order one in lattice units in a cross-over region that is indicated with a dashed line in Fig.~\ref{CleanPD}. 

The Fermi surface gets reconstructed at the O(2) Quantum Critical Point (QCP) between the SDW and the spin nematic phase. Inside the SDW, the spin rotation and translation symmetry breaking lead to a small Fermi surface. On the other hand, in the translationally-invariant spin nematic phase a simple mean-field argument shows that there is a large Fermi surface with a volume dictated by the standard Luttinger theorem, essentially because the electrons couple to $\Phi$, and $\langle \Phi\rangle = 0$ for the spin nematic. This mean-field argument is confirmed with quantum Monte Carlo simulations in the appendix. Because of the Fermi surface reconstruction, the O(2) QCP does not belong to the conventional XY universality class. The O(3) QCP, on the other hand, does fall into the standard Wilson-Fisher universality class.

\section{Effects of quenched potential disorder}\label{sec:dis}

Next we consider what happens to the phase diagram of the clean model upon adding potential disorder. Concretely, we will now consider following theory:
\begin{equation}
S = S_{\text{clean}} + S_{\text{dis}}\,,
\end{equation}
where $S_{\text{dis}}$ contains conventional doping-induced potential disorder:
\begin{equation}
S_{\text{dis}} = -\int\mathrm{d}\tau\, \sum_\r \mu(\r) \bar{\psi}(\r)\psi(\r) 
\end{equation}
We will take the random variable $\mu(\r)$ to be a simple white noise term
\begin{eqnarray}
\overline{\mu(\r)} & = & 0 \\
\overline{\mu(\r)\mu(\r')} & = & \sigma_\mu^2\, \delta_{\r,\r'}
\end{eqnarray}
Integrating out (high-energy) electrons induces following term for the bosonic field
\begin{equation}\label{disPhi}
S_{\text{dis}}^\Phi = -\int\mathrm{d}\tau\, \sum_\r \left( w^a(\r) \Phi_{aj}(\r) \right) \left( w^b(\r) \Phi_{bj}(\r) \right)\,,
\end{equation}
where $w^a(\r)$ are random variables. The $w^a(\r)$ are completely determined by $\mu(\r)$ (and other parameters in the fermion Hamiltonian), and in general will have decaying spatial correlations. However, for simplicity we consider them to also be white noise variables satisfying
\begin{eqnarray}
\overline{w^a(\r)} & = & 0 \\
\overline{w^a(\r)w^b(\r')} & = & \sigma_w^2\, \delta_{ab}\delta_{\r,\r'}
\end{eqnarray}
Note that $\sigma_\mu^2$ has units of energy squared, whereas $\sigma_w^2$ has units of energy. Physically, the term in Eq.~\eqref{disPhi} represents a coupling between the potential disorder and the $2Q$ CDW order induced by $\Phi$. So it is a random-field term for the composite (or vestigial) order parameter $\Phi\Phi^T$ (or $\vec{S}_Q \cdot \vec{S}_Q$).

\subsection{Effect of disorder on the spin nematic}
To understand the effect of the random-field term in Eq.~\eqref{disPhi} on the spin nematic phase, we first rewrite the spatial stiffness terms in $S_{\Phi}$ using $\vec{n}(\r)=(\cos\theta_\r,\sin\theta_\r)$ and $\vec{w}(\r) = |\vec{w}(\r)|(\cos\alpha_\r,\sin\alpha_\r)$. Up to constants we find
\begin{align}\label{spatialstiff}
& - J \sum_{\langle \r \r'\rangle}  \left[\vec{m}_\r\cdot\vec{m}_{\r'}\right] \cos\left(\theta_\r - \theta_{\r'}\right) \\
 & - \tilde{J} \sum_{\langle \r \r'\rangle} \left[\vec{m}_\r\cdot \vec{m}_{\r'}\right]^2 - \frac{1}{2}\sum_\r \vec{w}^2(\r) \cos(2\theta_\r - 2\alpha_\r) \,,\nonumber
\end{align}
where we have also included the random-field term. In the spin nematic phase there is no intrinsic CDW order, and hence there is a strong local pinning of the charge density to the random potential. So to capture the correct physics we can consider the situation $|\vec{w}(\r)|\gg J$, and take
\begin{equation}\label{thetar}
\theta_\r = \alpha_\r + \frac{1-\eta_\r}{2}\pi\,,
\end{equation}
with $\eta_\r = \pm 1$. The spin nematic phase is characterized by a non-zero spin stiffness, and hence we consider the case where also $\tilde{J}\gg J$. Low-energy configurations of the $\tilde{J}$ term are given by
\begin{eqnarray}\label{mr}
\vec{m}_\r & = & \vec{N}_\r s_\r\,,
\end{eqnarray}
where $s_\r = \pm 1$, and $\vec{N}_\r$ varies slowly on the lattice scale. Plugging Eqs.~\eqref{thetar} and~\eqref{mr} in Eq. \eqref{spatialstiff} we obtain (again up to a constant)
\begin{equation}
-  \sum_{\langle \r \r'\rangle} J_{\r,\r'}  \left[\vec{N}_\r\cdot\vec{N}_{\r'}\right] S_\r S_\r'  - \tilde{J} \sum_{\langle \r \r'\rangle} \left[\vec{N}_\r\cdot \vec{N}_{\r'}\right]^2
\end{equation}
Here we have defined $S_\r = s_\r\eta_\r=\pm 1$. Note that both $s_\r$ and $\eta_\r$ transform non-trivially under the $\mathbb{Z}_2$ gauge redundancy, but $S_\r$ is gauge invariant and hence physical. In the above equation we have also introduced
\begin{equation}
J_{\r,\r'} = J \cos\left(\alpha_\r - \alpha_{\r'}\right)
\end{equation}
Because $\vec{N}_\r$ is a smoothly varying field we can approximate $\vec{N}_\r\cdot \vec{N}_{\r'}\sim 1$, and obtain
\begin{equation}
-\sum_{\langle  \r \r'\rangle} J_{\r,\r'} S_\r S_{\r'}
\end{equation}
As the average value of $J_{\r,\r'}$ is zero, this is a classical Ising spin glass Hamiltonian for the variables $S_\r$. 

To see whether quantum fluctuations destroy the potential spin glass order at zero temperature, we consider the temporal gradient term for $\Phi$. In the path integral, the Ising variables $S_\r$ change sign along the time direction when $\theta_\r$ changes by $\pi$, or when $\vec{N}_\r$ rotates to $-\vec{N}_\r$. Again assuming that in the spin nematic phase the fluctuations in $\vec{N}_\r$ are subdominant, we write 
\begin{equation}
\frac{J}{v^2} (\partial_\tau \Phi)^2 \sim \frac{J}{v^2} (\partial_\tau \theta)^2
\end{equation}
Using this term we obtain the following on-site equation of motion for $\theta_\r$:
\begin{equation}
\frac{\mathrm{d}^2\theta_\r}{\mathrm{d}\tau^2}=\frac{v^2\vec{w}^2(\r)}{2J} \sin(2\theta_\r-2\alpha_\r) 
\end{equation}
We now consider an instanton solution to this equation of motion, representing a tunneling process where at time $\tau_0$, $\theta_\r$ changes by $\pi$:
\begin{equation}
\theta_\r(\tau) = \alpha_\r + \frac{\pi}{2} + \text{arcsin}\,\text{tanh}\left(\frac{v|\vec{w}(\r)|}{\sqrt{J}}(\tau-\tau_0)\right) 
\end{equation}
The difference between the instanton action and the minimal action is given by  
\begin{equation}
S_{\text{inst}} - S_{\text{min}} = \frac{4 |\vec{w}(\r)|\sqrt{J}}{v} 
\end{equation}
From this expression we can estimate the strength of the transverse field term in the quantum Ising Hamiltonian for $S_\r$:
\begin{equation}
h_\r \sim \frac{v^2}{J}e^{-\frac{4 \sqrt{J}}{v}|\vec{w}(\r)|}\label{hr}
\end{equation}
We see that $h_\r$ decays exponentially with the strength of the potential disorder, which is the familiar exponential decay of a tunneling amplitude with the barrier height. We thus conclude that even though $\theta_\r$ was quantum disordered in the spin nematic phase of the clean model, potential disorder efficiently suppresses quantum fluctuations in $S_\r$ such that there can nevertheless be zero-temperature Ising spin glass order.

\begin{figure*}
  \includegraphics[scale=0.6]{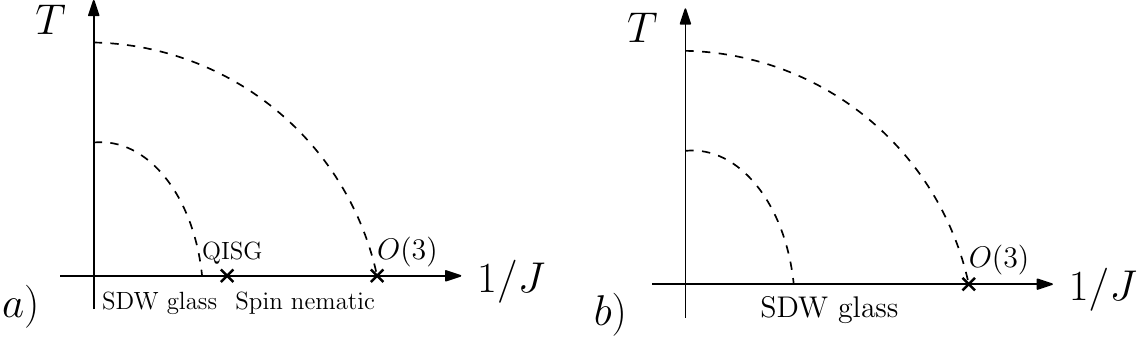}
  \caption{Phase diagrams of the disordered model. In (a) we consider the weak disorder case, with a zero-temperature Quantum Ising Spin Glass (QISG) transition between the SDW glass and the spin nematic. (b) corresponds to the stronger-disorder case, where the SDW glass order extends all the way to the O(3) transition of the clean model. In both phase diagrams, the dashed lines represent cross-over regions. The inner dashed line is where $\xi_V$ (defined around Eq.~\eqref{xiV}) becomes order one in lattice units. The outer dashed line, which is the same as in the clean model, corresponds to the onset of local spin nematic order. Potential Quantum Griffiths phases in the vicinity of the QCPs are not shown. As explained in the text, quantum phase transitions can also be rounded by random-mass-type disorder.}
  \label{DisPD}
\end{figure*}

Let us now explain the physical origin of the Ising spin glass behavior. We start from the fact that the potential disorder generates a particular inhomogeneous charge distribution. The spin nematic, properly thought of as a fluctuating SDW, will pin a static spin configuration to this charge configuration. This happens in such a way that troughs in the charge density, corresponding to hole-rich regions, align with spin-flip domain walls in the anti-ferromagnetic (AFM) pattern. However, the randomly pinned CDW configuration contains dislocations where such troughs end. But since the AFM domain walls are necessarily closed, it is not possible to align domain walls with all the troughs. This is the origin of the frustration in the Ising spin glass model. 

One mechanism to relieve the frustration is to allow for point singularities in $\vec{N}_\r$ around which the the spin direction changes sign (up to now we have by hand excluded such singular configurations of $\vec{N}_\r$). In particular, by aligning a singularity in $\vec{N}_\r$ with every dislocation of the pinned CDW the frustration is completely resolved. This is because now every charge dislocation realizes a half-vortex of the SDW order. 

Above, we have assumed that introducing a finite density of half-vortices in $\Phi$ costs more energy than introducing a finite density of (short) $\mathbb{Z}_2$ domain walls where $\Phi$ changes sign. Within the effective theory this was justified by $\tilde{J}>J$, but it also seems physically reasonable from a microscopic perspective, as large parts of the $\mathbb{Z}_2$ domain walls can be realized as charge troughs where the AFM order goes to zero, but does \emph{not} flip its direction. These parts of the domain walls thus lie entirely inside the hole-rich regions where the AFM magnitude is small, leading to a small energy cost. On the other hand, introducing a singularity in $\vec{N}_\r$ requires (1) the AFM order to vanish near the core, which includes regions with low hole density, and (2) the spin direction to twist in regions where the AFM order is maximal. These effects lead to a large core energy and large elastic energy for the half-vortices.

\subsection{Effect of disorder on the SDW}
Next we consider the effect of potential disorder on the SDW phase in the clean model. We focus on the region near the SDW-spin nematic phase transition, where the spin stiffness $K_s \sim  \langle \Phi\rangle^2 J +2|Q|^2\tilde{J}$ is guaranteed to be much larger than the charge stiffness $K_c \sim \langle \Phi\rangle^2 J$ (in these expressions, expectation values are taken at zero temperature). This case was analyzed by Mross and Senthil in Ref.~\cite{Mross2015}. Building on earlier work on disordered vortex lattices~\cite{Giamarchi1994,Giamarchi1995,LeDoussal2000,Gingras1996}, these authors found that in the weak disorder case there is an important emergent length scale 
\begin{equation}\label{xiV}
\xi_V \sim \xi_L e^{\sqrt{\alpha\ln\xi_L}}\,,
\end{equation}
with $\alpha$ a function of temperature, and $\xi_L = K_c/\sigma_w^2$ the Larkin length~\cite{Larkin1970}. At distances smaller than $\xi_V$, the SDW behave like a `Bragg glass'~\cite{Giamarchi1994,Giamarchi1995,LeDoussal2000} with algebraic order in $\Phi$. At larger distances $K_c$ renormalizes to zero and the system behaves essentially like the disordered spin nematic discussed above. This state was called a SDW glass by Mross and Senthil~\cite{Mross2015}.

\subsection{Phase diagram with disorder}
Putting everything together we arrive at the two phase diagrams for the disordered model shown in Fig.~\ref{DisPD} (in these phase diagrams we do not show any potential Quantum Griffiths phases in the vicinity of the QCPs). 

In Fig.~\ref{DisPD}(a) we consider the weak-disorder case. This leads to a phase diagram where the O(2) QCP and the associated BKT line are replaced with a cross-over region where $\xi_V$ becomes order one in lattice units. As the disorder is weak, upon decreasing $J$ quantum fluctuations destroy the Ising spin glass order for the $S_\r$ variables. This leads to a Quantum Ising Spin Glass (QISG) transition between the SDW glass and the spin nematic. Due to the coupling of the Ising variables to the Fermi surface and the associated Landau damping the sharp QISG transition is most likely rounded by the random-mass-type disorder in Eq.~\eqref{hr}~\cite{Vojta2003}.

Both the SDW glass and spin nematic order disappear at non-zero temperature. The cross-over region associated with the onset of local spin nematic order survives from the clean model. At smaller $J$, the O(3) QCP between the spin nematic and the symmetric phase remains, and is in the Wilson-Fisher universality class just as in the clean model.

In Fig.~\ref{DisPD}(b) we consider the case with stronger disorder. The cross-over regions are the same as before. However, with stronger disorder the QISG transition gets pushed to smaller values of $J$. We consider the case where the disorder is sufficiently strong that the QISG transition merges with the O(3) QCP of the clean model. Now the O(3) transition separates the SDW glass from the symmetric phase, and it no longer belongs to the Wilson-Fisher universality class. To see this, note that the effective theory for the O(3) QCP in this second (stronger-disorder) scenario is
\begin{eqnarray}\label{SSM}
 S & = & \int \mathrm{d}\tau \sum_{\k} \bar{\psi}(\k)(\partial_\tau + \varepsilon_\k)\psi(\k) - \sum_\r \mu(\r)\bar{\psi}\psi \nonumber \\
 & + & \int \mathrm{d}\tau \sum_{\r}g(\r)  \, (-1)^{r_x+r_y} \vec{N}\cdot \bar{\psi}\vec{\sigma} \psi  \nonumber \\
 & + & \tilde{J}\int\mathrm{d}\tau\,\sum_{\r} \frac{1}{\tilde{v}^2} \left(\partial_\tau \vec{N}\right)^2 + \left(\nabla_j \vec{N}\right)^2\,,
\end{eqnarray}
where $\vec{N}$ is tuned to criticality, and we have defined
\begin{equation}
g(\r) =  g\, S_\r\cos(Qr_x - \alpha_\r)
\end{equation}
At zero temperature, one can replace $S_\r$ by its expectation value due to the spin glass order. Crucially, we see that the critical field $\vec{N}$ is coupled to the Fermi surface, and hence this theory realizes a 2+1-dimensional quantum critical metal~\cite{Hertz1976,Moriya,Millis1993}, which flows to strong electron-boson coupling~\cite{SachdevBook}. The theory in Eq.~\eqref{SSM} is exactly the universal theory for strange metals put forward in Refs.~\cite{Aldape2022,Esterlis2021,Guo2022,Patel2023}. Interestingly, the disordered SDW model naturally realizes a situation in which the average value of $g(\r)$ is zero, a property which was argued to lead to a Planckian scattering rate at the quantum critical point of $\vec{N}$~\cite{Aldape2022,Esterlis2021,Guo2022,Patel2023}. As the two-dimensional Ising spin glass order disappears at non-zero temperature, thermal fluctuations automatically lead to a self-averaging of $S_\r$, and hence the sign of $g(\r)$.

The model in Eq.~\eqref{SSM} was first studied in a large-$N$ limit, and the corresponding saddle-point solutions were shown to exhibit marginal Fermi liquid behaviour~\cite{Varma1989,Ruckenstein1991} and linear-T resistivity~\cite{Aldape2022,Esterlis2021,Guo2022,Patel2023}. Recent numerical works~\cite{Patel2024_2,Patel2024} have gone beyond the large-$N$ approximation and found that the O(3) QCP is rounded by the disorder, and is replaced by a gapless Quantum Griffiths phase where linear-T resistivity is realized over an extended parameter region~\cite{Patel2024}.

\begin{figure*}
    a)
    \includegraphics[scale=0.25]{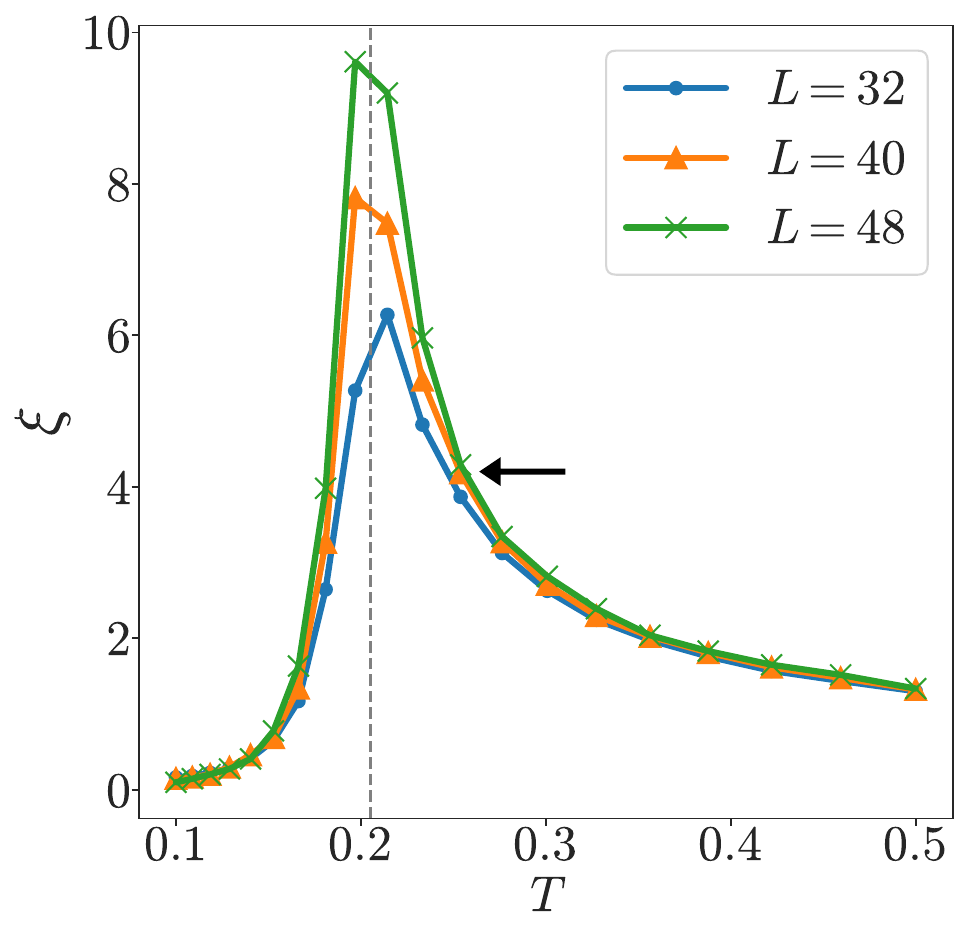}
    b) 
    \includegraphics[scale=0.28]{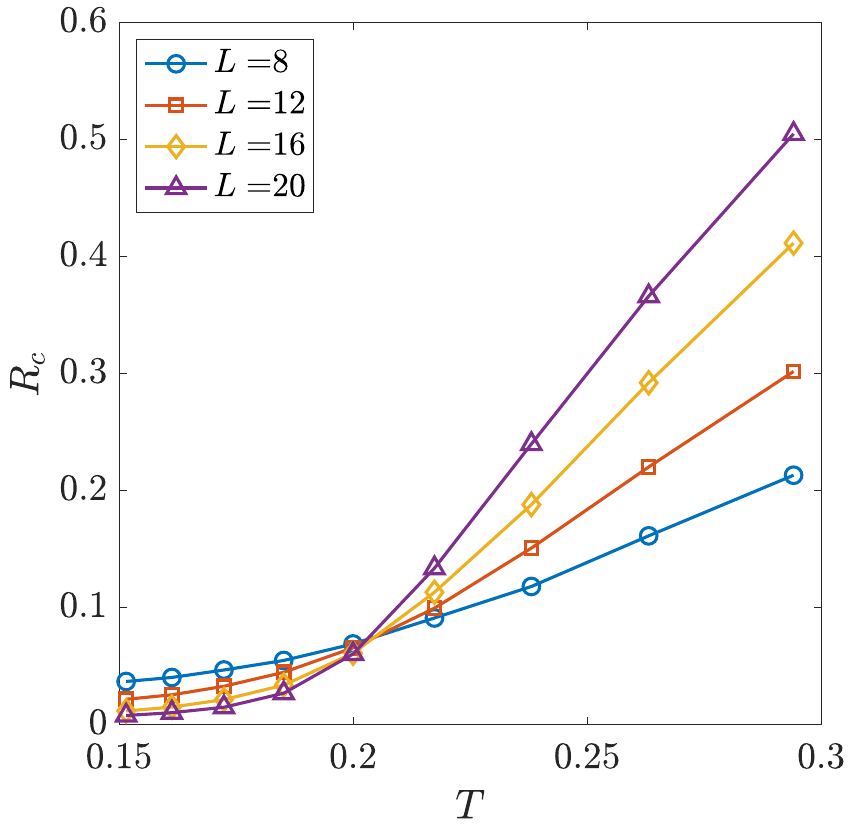}
    c) 
    \includegraphics[scale=0.28]{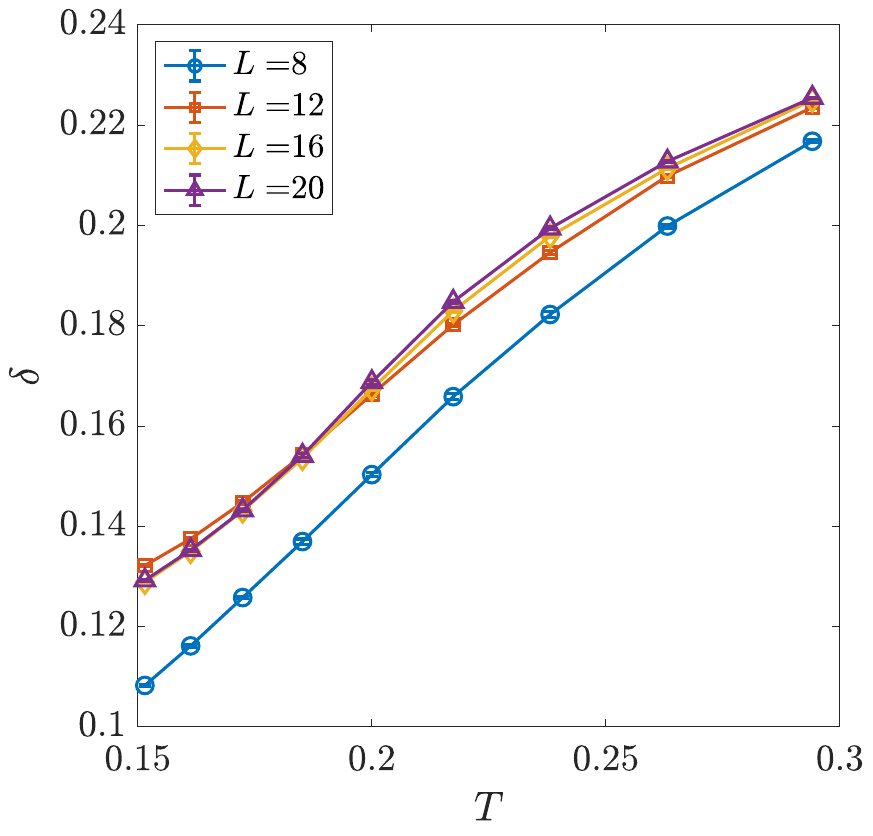} \\
    \vspace{0.5 cm}
    d)
    \includegraphics[scale=0.3]{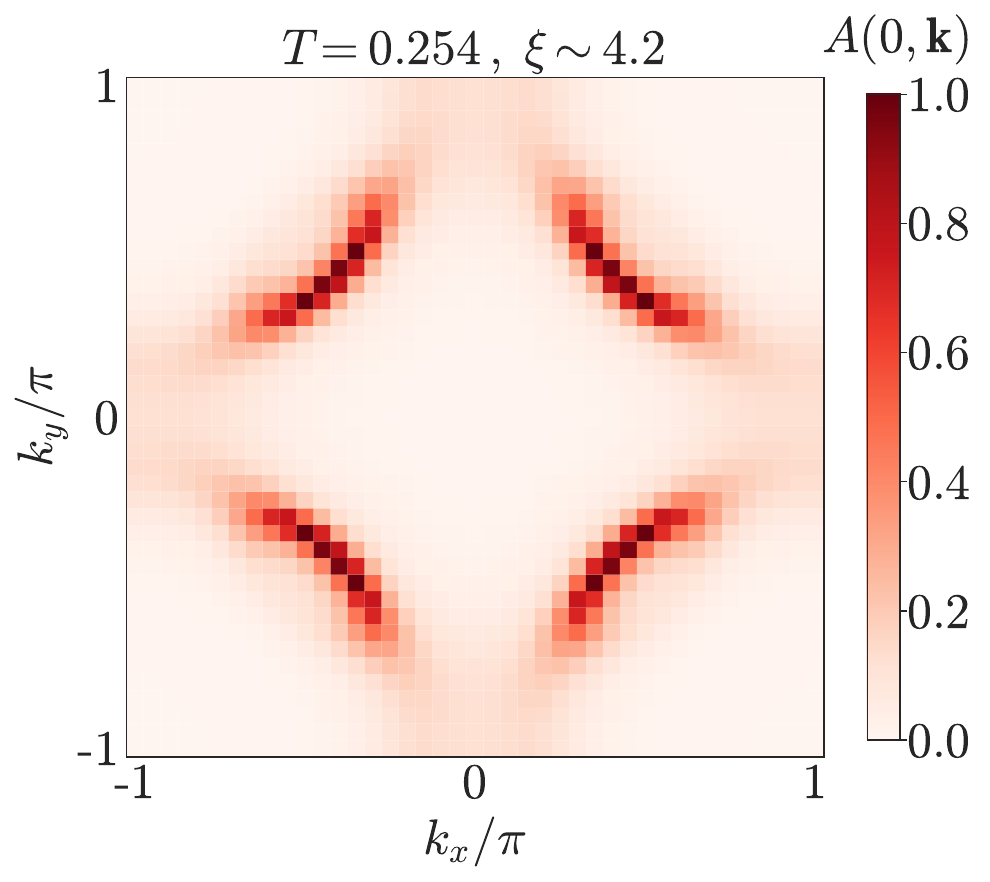}
    e)
    \includegraphics[scale=0.3]{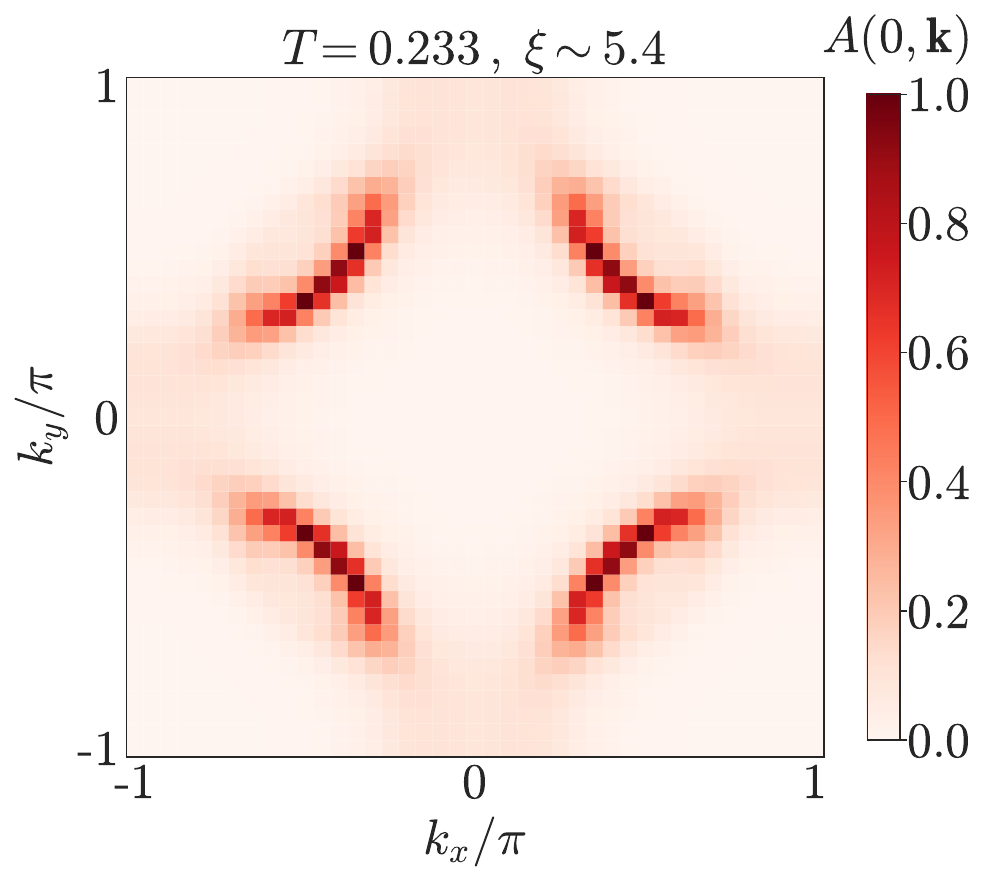}
    \caption{Numerical results on the plaquette-Ising model in Eq.~\eqref{HIP}. (a) AFM correlation length as a function of temperature. (b) Correlation ratio as a function of temperature. (c) Hole doping as a function of temperature. (d) Normalized zero-frequency spectral weight at $T=0.254$. (e) Normalized zero-frequency spectral weight at $T=0.233$. (d) and (e) are obtained with $L=40$.}
    \label{fig:plaqIsing}
\end{figure*}

\section{Thermal fluctuations and Fermi arcs}\label{sec:arcs}

In this section we focus on the thermally disordered SDW glass, i.e. we are interested in the situation where thermal fluctuations have washed out the glassy order (by assumption, we stay away from the QCP discussed in the previous section, so quantum critical fluctuations and strange metallic behaviour are absent). We ask the question whether pseudo-gap-like physics can emerge in this region of the phase diagram, in a way that is consistent with experiment.

As the SDW glass has local anti-ferromagnetic (AFM) order, let us first recall that the zero-frequency spectral weight of a hole-doped AFM on the square lattice shows 4 small hole pockets centered at $(\pm \pi/2,\pm \pi/2)$. The `fronts' of these pockets, i.e. the sides facing the $\Gamma$ point, have a large spectral weight, whereas the backsides facing the $X$-point have a significantly smaller spectral weight. Because of these properties, the zero-frequency spectral weight of a hole-doped AFM looks similar to the Fermi arcs of the pseudogap metal. However, in experiment only a short AFM correlation length is found at the doping levels corresponding to the pseudogap region. So, in order to interpret the pseudogap as a fluctuating hole-doped AFM, one has to answer (at least) two questions: (1) if at the temperatures of the pseudogap there is fluctuating AFM order, why does no long-range AFM order develop at low temperatures and why is there no divergence in the AFM correlation length as the system is cooled?, and (2) is a short AFM correlation length actually sufficient to obtain bright Fermi arcs? The SDW glass provides an answer to the first question as it has no long-range AFM order, and upon cooling the AFM correlation length saturates at a finite value determined by the size of the AFM clusters in the glassy order. In the remainder of this section we address the second question and show, as a proof-of-principle, that it is indeed possible to obtain bright Fermi arcs in a thermally fluctuating short-range AFM, provided that the thermal fluctuations are of the Ising-type, as is the case in the SDW glass.

Our approach is to simulate a phenomenological model with Monte Carlo, and assume that the spin nematic correlation length is larger than the system sizes used in the numerics. This allows us to consider easy-axis spin degrees of freedom, i.e. an Ising spin system. We think of the Ising degrees of freedom as the $S_{\r}$ variables discussed above. However, we will not explicitly consider the disorder in this section. Instead, we work with a uniform Ising Hamiltonian for $S_\r$, as we expect that as far as the fermions are concerned, a background consisting of a thermally disordered clean Ising system should behave similarly as a thermally disordered Ising spin glass, even though the former orders at $T_c>0$ while the latter only orders at $T_c=0$. We thus use the clean system at $T>T_c$ to emulate the physics of the Ising spin glass at $T>0$. The reason for this is simply that simulating a clean system can be done with significantly fewer computational resources, allowing us to go to larger system sizes and hence achieve a better momentum resolution for the electronic spectral weight.

The model we consider in this section has spinful fermions hopping on a $L\times L$ square lattice with periodic boundary conditions, labeled by coordinates $\r$. On one quarter of the plaquettes of the fermion lattice we place a classical Ising spin. The Ising spins are indexed by coordinates $\R$, which live on a $L/2\times L/2$ lattice. The connection between the two sets of coordinates is $\r = 2\R + (m,n)$, where $m,n\in \{0,1\}$. The specific Hamiltonian we consider is
\begin{eqnarray}
H & = & -t \sum_{\langle \r\r'\rangle} c^\dagger_{\r}c_{\r'} - t'\sum_{\langle\langle \r\r'\rangle\rangle} c^\dagger_{\r}c_{\r'} + h.c.  -\mu \sum_\r c^\dagger_\r c_\r \nonumber\\
& + & g\sum_\r (-1)^{r_x+r_y} S_{\R(\r)} c^\dagger_\r \sigma^z c_\r - \sum_{\langle \R\R'\rangle} J S_\R S_{\R'} \label{HIP}\,,
\end{eqnarray}
where $\R(\r)$ denotes the Ising spin lattice site associated with the fermion lattice site $\r$, using the relation between the $\R$ and $\r$ coordinates given above. This model can be straightforwardly simulated with classical Monte Carlo for the Ising spins. In our numerics we update the spins by constructing Wolff or Swendsen-Wang clusters in the usual way, and then flip the cluster with a probability $\text{min}(1,e^{-\beta \Delta F})$, where $\Delta F$ is the change in free energy for the fermions. We also use parallel tempering to further reduce autocorrelation times. In all the numerical simulations presented in the main text we use the parameters $t=1$, $t'=-0.35$, $\mu = -1.4$, $g=1$, and $J=0.1$.

An important quantity we measure is the spin correlation function of the fermions:
\begin{equation}
C(\r) = \langle c^\dagger_\r \sigma^z c_\r c^\dagger_0 \sigma^z c_0\rangle
\end{equation}
We will denote the Fourier transform of this correlation function as $\chi(\q)$. Near $\Q = (\pi,\pi)$, $\chi(\q)$ will take on following form:
\begin{equation}
\chi(\Q+\q) \sim L^2m^2\delta_{\q,0}+\frac{\chi_0}{(\xi\q)^2+1}\,, \label{chiQq}
\end{equation}
where $m$ is the magnetization, and $\xi$ is the anti-ferromagnetic correlation length. We extract $\xi$ from Eq. \eqref{chiQq} by fitting a Lorentzian to $\chi(\Q+\q)$. For this fit we exclude $\q=0$, and use $|\q|\leq 5\times 2\pi/L$. We note that Eq. \eqref{chiQq} needs to be modified very close to the critical point. Nevertheless, the above procedure still allows us to extract a length scale for the anti-ferromagnetic correlations. This length scale will deviate slightly from the actual correlation length close to the critical point, but for our purposes this is not too important. From $\chi(\q)$ we also obtain the correlation ratio
\begin{equation}
R_c = \frac{\chi(\Q + \Delta\q)}{\chi(\Q)}\,,
\end{equation}
with $\Delta\q= (\frac{2\pi}{L},0)$. At the critical point, $R_c$ will (to leading order) be independent of the system size $L$~\cite{Ribhu2015,Ribhu2016}.

In Figs.~\ref{fig:plaqIsing} (a) and (b) we show $\xi$ and $R_c$ as a function of temperature for different system sizes. We can identify a clear peak in the correlation length, growing with $L$, at a temperature where also the $R_c(T)$ curves obtained at different $L$ cross. This allows us to estimate the critical point as $T_c \sim 0.205$ (indicated with a dashed gray line in Fig.~\ref{fig:plaqIsing}(a)). Note that this value is lower than the critical temperature for the decoupled Ising model, which here would be $\sim 0.227$. We can understand this shift in $T_c$ as a consequence of the fermion-induced Ruderman-Kittel-Kasuya-Yosida (RKKY) interaction between the Ising spins, as explained in detail in the appendix.

We also measured the filling $n = \langle c^\dagger_\r c_\r \rangle$. Note that without coupling to the Ising spins, the van Hove singularity of the free-electron band would be on the Fermi surface. However, the interaction with the Ising spins significantly shifts the filling. To understand why, consider a situation in which the Ising spins are ordered, and the anti-ferromagnetism induces a bandgap (not necessarily at the Fermi energy). The chemical potential $\mu$ inducing van Hove filling for decoupled electrons now leads to a completely different filling, due to the strong band reconstruction. Because of this effect we expect $n$ to depend strongly on temperature. This is confirmed in Fig.~\ref{fig:plaqIsing} (c), where we plot the hole density $\delta$, defined as $n = 1-\delta$. We are mostly interested in the temperature range $T\sim 0.21 - 0.25$, for which $\delta \sim 0.17 - 0.21$.

Finally, we also measured the fermion spectral weight
\begin{equation}
A(\omega,\k) = -\frac{1}{\pi}\text{Im}\,G(\omega+i\epsilon,\k)\,,
\end{equation}
where $G(\omega,\k)$ is the fermion Green's function, and we used $\epsilon = 0.01$. In Fig.~\ref{fig:plaqIsing} (d) we show the normalized zero-frequency spectral weight obtained at $T\sim 0.254$ and $L=40$, for which we find an anti-ferromagnetic correlation length $\xi \sim 4.18$ (the point in Fig.~\ref{fig:plaqIsing} (a) that goes with Fig.~\ref{fig:plaqIsing} (d) is indicated with a black arrow). We expect that this estimate for $\xi$ is quite reliable, as (1) at this temperature $\xi$ is converged with system size, and (2) the value obtained from numerics agrees remarkably well with the expected relation 
\begin{equation}\label{xiprop}
\xi \propto \left(\frac{T-T_c}{T_c} \right)^{-1}\,,
\end{equation}
if we simply use a proportionality constant of one. From Fig.~\ref{fig:plaqIsing} (d) see that even with this short-range AFM, the zero-frequency spectral weight is strongly suppressed in the anti-nodal regions -- leading to four bright Fermi arcs in the nodal regions. To be quantitative, we find that the maximal spectral weight in the anti-nodal region (defined as the maximal spectral weight along the cut $\k = (k_x,\pi)$) is suppressed by a factor of $\sim 7.2$ compared to the overall maximal spectral weight. 

In Fig.~\ref{fig:plaqIsing} we also show the normalized zero-frequency spectral weight obtained at $T\sim 0.233$, for which we find $\xi \sim 5.4$ at $L=40$. In this case, the spectral weight in the anti-nodal region is suppressed by a factor $\sim 9.5$ compared to the maximal spectral weight. However, for this temperature we find that $\xi$ is not yet converged with $L$. Using Eq.~\eqref{xiprop}, again with proportionality constant one, we obtain an estimate of $\xi \sim 7.3$ in the thermodynamic limit. We note that larger $L$ will also lead to a higher momentum resolution, which we expect will lead to an increase of the maximal spectral weight in the nodal region, where $A(0,\k)$ varies rapidly with $\k$. So in the thermodynamic limit we expect that the relative suppression of the spectral weight in the anti-nodal region will be larger than $9.5$.

\section{Discussion}\label{sec:disc}
The main goal of this paper has been to make two observations:
\begin{enumerate}
\item Adding potential disorder to a model of fluctuating spin stripes naturally produces the theory of strange metals put forward in  Refs.~\cite{Aldape2022,Esterlis2021,Guo2022,Patel2023}.
\item A simple (and numerically solvable) model for thermal fluctuations in a SDW glass produces clear Fermi arcs in the zero-frequency spectral weight already with short-range AFM correlations.\vspace{0.5 cm}
\end{enumerate}
The bearing that these observations have on the actual physics of the cuprate superconductors is at present unclear and requires further study. Nevertheless, we believe that the phase diagram of disordered spin stripes -- which contains both the strange metal theory and the Ising variables of the numerically solvable model with Fermi arcs -- seems to naturally fit together a compelling number of experimental and theoretical pieces of the puzzle.

Finally, we end with some open questions for future work. First, the Ising spin glass found here introduces effective two-level systems~\cite{Anderson1972,Phillips1972}. It was shown that coupling fermions to these two-level systems induces a stable non-Fermi liquid phase~\cite{Tulipman2024} (not necessarily with Marginal Fermi liquid behaviour, but instead with a continuously-varying frequency exponent in the fermion self-energy). It would therefore be interesting to try and incorporate these two-level systems explicitly into the theory studied in this work. Secondly, we note that electron-boson models of the type in Eq.~\eqref{SSM} have also been shown to reproduce the linear-in-$B$ scaling of the resistivity, with $B$ an out-of-plane magnetic field~\cite{Kim2024,Kim2025}. In these works, glassy charge-density-wave order --a natural side-product of the SDW glass-- plays an important role. It would thus be interesting to recover the linear-in-$B$ scaling within the disordered spin density wave theory.

\color{black}

\begin{acknowledgements}
This research was supported by the European Research Council under the European Union Horizon 2020 Research and Innovation Programme via Grant Agreement No. 101076597-SIESS (X.Z. and N.B.), and by a grant from the Simons Foundation (SFI-MPS-NFS-00006741-04) (N.B.).
\end{acknowledgements}

\bibliography{bib}

\clearpage
\appendix

\onecolumngrid

\section{The $\frac{O(3)\times O(2)}{\mathbb{Z}_2}$ boson model}
In this appendix, we present a more general effective model for fluctuating spin stripes, and numerically determine the phase diagram. In the next appendix we couple this bosonic model to a Fermi surface, in such a way that the resulting Euclidean action is amenable to sign problem-free Quantum Monte Carlo (QMC) simulations.

\subsection{Introduction of the model}
In the appendix we work with the three-component complex field $\phi$, which is related to the $2\times 3$ real matrix field $\Phi$ of the main text as $\phi_\eta = \Phi_{1,\eta}+i\Phi_{2,\eta}$. The purely bosonic part of the action is given by
\begin{eqnarray}
	S_B&=&\int d\tau [\frac{1}{2}\sum_{\r;\eta}J|\frac{1}{c}\partial_\tau\phi_{\eta}(\r,\tau)|^2+J|\boldsymbol{\nabla}_r \phi_{\eta}(\r,\tau)|^2-m|\phi_{\eta}(\r,\tau)|^2 + \lambda \sum_{\r}(\sum_{\eta}|\phi_{\eta}(\r,\tau)|^2)^2-v \sum_{\r}|\sum_{\eta}(\phi_{\eta}(\r,\tau))^2|^2	\nonumber\\
	&-&J_v\sum_{\r}\sum_{dr=dx,dy}(\frac{\sum_{\eta}(\phi_{\eta}(\r+dr,\tau))^2\sum_{\nu}(\phi_{\nu}^*(\r,\tau))^2}{|\sum_{\eta}(\phi_{\eta}(\r+dr,\tau))^2\sum_{\nu}(\phi_{\nu}^*(\r,\tau))^2|}+c.c.)-J_u\frac{2\Tr[Q(\r+dr,\tau)Q(\r,\tau)]}{\sqrt{\Tr[(Q(\r+dr,\tau))^2]\Tr[(Q(\r,\tau))^2]}}] \nonumber\\
	&=&\int d\tau [-\frac{J}{c^2\Delta\tau^2}\sum_{\r;\eta}(\Re[\phi_{\eta}(\r,\tau+\Delta\tau)]\Re[\phi_{\eta}(\r,\tau)]+\Im[\phi_{\eta}(\r,\tau+\Delta\tau)]\Im[\phi_{\eta}(\r,\tau)]) \nonumber\\
	&-&\sum_{\r}\sum_{dr=dx,dy}\sum_{\eta}J(\Re[\phi_{\eta}(\r+dr,\tau)]\Re[\phi_{\eta}(\r,\tau)]+\Im[\phi_{\eta}(\r+dr,\tau)]\Im[\phi_{\eta}(\r,\tau)])-m'\sum_{\r}\sum_{\eta}|\phi_{\eta}(\r,\tau)|^2 \nonumber\\
	&+& \lambda \sum_{\r}(\sum_{\eta}|\phi_{\eta}(\r,\tau)|^2)^2-v\sum_{\r}|\sum_{\eta}(\phi_{\eta}(\r,\tau))^2|^2 \nonumber\\
	&-&J_v\sum_{\r}\sum_{dr=dx,dy}(\frac{\sum_{\eta}(\phi_{\eta}(\r+dr,\tau))^2\sum_{\nu}(\phi_{\nu}^*(\r,\tau))^2}{|\sum_{\eta}(\phi_{\eta}(\r+dr,\tau))^2\sum_{\nu}(\phi_{\nu}^*(\r,\tau))^2|}+c.c.)-J_u\frac{2\Tr[Q(\r+dr,\tau)Q(\r,\tau)]}{\sqrt{\Tr[(Q(\r+dr,\tau))^2]\Tr[(Q(\r,\tau))^2]}}]\label{SB}
\end{eqnarray}
Here, the $J,m$ and $\lambda$ terms form the standard O(6) $\phi^4$ theory, while the $v$ term breaks the O(6) symmetry down to O$(3)\times$O(2) and prefers the phases of the three complex components of $\phi$ to be identical, so that the potential is minimized by $\phi =  Ae^{i\theta}$, where $A$ is real. As in the main text, the $v$ term thus favors spin density wave (SDW) states over spin spiral states. In Eq.~\eqref{SB} we have also defined 
\begin{equation}
	Q_{\eta,\nu}(\r,\tau)=\Re[\phi_{\eta}(\r,\tau)\phi_{\nu}^*(\r,\tau)]-\frac{1}{3}\delta_{\eta,\nu}\sum_{\eta}|\phi_{\eta}(\r,\tau)|^2\,,
\end{equation}
which, similarly as in the main text, is the spin nematic order parameter. The $J_u$ and $J_v$ terms are stiffness terms for respectively the composite spin nematic and the composite charge density wave order parameters. Without these terms, the velocities of the O(3) and O(2) Goldstone modes of the SDW are identical, pointing to an accidental O(4) symmetry. Both the $J_u$ and $J_v$ term lift this degeneracy.

Below we will take $m$ and $\lambda$, the coefficients of the field potential, to be quite large ($m,\lambda\approx 10$). Also note that we take the bosonic mass term to be negative. As a result, the bosonic field predominantly fluctuates at the bottom of a deep potential well, which is located at a non-zero value for $|\phi|$. The denominators in the $J_v$ and $J_u$ terms are a normalization factor, which ensure that the effective magnitudes of these terms are indeed $J_v$ and $J_u$, and not $|\phi|^4\times J_v$ or $|\phi|^4\times J_u$. Because the denominators are analytic away from $|\phi|=0$, one can expand them around the minimum of the field potential. The additional terms that are generated by this expansion involve the amplitude or Higgs mode, which has a large mass and hence is not important for the low-energy physics. Our choice to make the bosonic fields fluctuate at the bottom of a deep potential well is to emulate the non-linear sigma model description of the Landau-Ginzburg theory, as adopted in Section~\ref{sec:clean} in the main text.

\subsection{Wolff cluster update}
For the QMC simulations we use Wolff cluster updates to reduce autocorrelation time and increase the sampling efficiency. However, the Wolff updates should be combined with local updates, since the former cannot update the longitudinal mode $|\phi|$. We divide the Wolff update in $O(3)$ and $O(2)$ parts, and discuss them separately.

The $O(3)$ Wolff update is similar to the update used for the conventional $O(3)$ model, but with the additional constraint that the real and imaginary $O(3)$ parts of $\phi$ should be reflected with the same mirror. Any such $O(3)$ update together with rotations will not affect the $J_v$ term, while the influence on the $J_u$ term should be identified carefully. It is easy to show that the same mirror flip for both $\phi_{\eta}(\r,\tau)$ and $\phi_{\eta}(\r+dr,\tau)$ within a cluster does not change the $J_u$ term:
\begin{equation}
	\Tr[Q(\r+dr,\tau)Q(\r,\tau)]=\Tr[R (Q(\r+dr,\tau)) R^{T} R (Q(\r,\tau)) R^{T}].
\end{equation}
The detailed balance condition is satisfied by adding a bond to the cluster with probability 
\begin{equation}
P_3(\r_0,\tau_0;\r,\tau)=1-e^{\min\{0,2J \phi^*_{\alpha}(\r_0,\tau_0)\phi_{\alpha}(\r,\tau)+4 z_u\Re[\phi^*(\r_0)\cdot \phi(\r) \phi^*_{\alpha}(\r)\phi_{\alpha}(\r_0)+\phi^*(\r_0)\cdot \phi(\r) \phi^*_{\alpha}(\r)\phi_{\alpha}(\r_0)-2|\phi^*_{\alpha}(\r)\phi_{\alpha}(\r_0)|^2]\}}
\end{equation}
Here, $z_u$ represents the coefficient before $\Tr[Q(\r+dr,\tau)Q(\r,\tau)]$, and $\phi_{\alpha}(\r_0,\tau_0)$ is the component of $\phi$ perpendicular to the $\alpha$ plane.

For the $O(2)$ update, it is obvious that it will not affect the $J_u$ term. And the $J_v$ term corresponds to $\cos[2\theta(\r)-2\theta(\r+dr)]$, while the $J$ term corresponds to $\cos[\theta(\r)-\theta(\r+dr)]$. After choosing a 2D mirror plane labeled with vector $(\cos(\alpha),\sin(\alpha))$ for reflecting $\theta$ which labels the phase between real and imaginary part of each $\phi$ components, we need to find the effective mirror for reflecting $2\theta$. This mirror should be $(\cos(2\alpha-\pi/2),\sin(2\alpha-\pi/2))$ and will also be used in the probability of including the next site in the $O(2)$ cluster
\begin{equation}
	P_2(\r_0,\tau_0;\r,\tau)=1-e^{\min\{0,2J \phi^*_{\alpha}(\r_0,\tau_0)\phi_{\alpha}(\r,\tau)+2 z_v (\phi^*(\r_0,\tau_0))_{2\alpha}^2 (\phi(\r,\tau))_{2\alpha}^2\}}.
\end{equation}
Here, the $z_v$ represents the coefficient before $\sum_{\eta}(\phi_{\eta}(\r+dr,\tau))^2\sum_{\nu}(\phi_{\nu}^*(\r,\tau))^2$ and $(\phi(\r_0,\tau_0))_{2\alpha}^2$ is the $(\phi(\r_0,\tau_0))^2$ field component perpendicular to the mirror plane reflecting $2\theta$.

To summarize, the update process is as follow. First, choose a site $(\r_0, \tau_0)$ and a mirror plane $\alpha$ randomly, reverse about the mirror plane $\alpha$ and add it to the cluster. Next, check the sites $(\r, \tau)$ connecting with $(\r_0, \tau_0)$, reverse and add it to the cluster with probability $P$. Then, if the site is accepted it will not be checked anymore. Repeat the last step until all the boundary sites of the cluster are checked and rejected.


\subsection{Phase diagram}
We set $\Delta\tau=0.1,m'=c=1/\Delta\tau,\lambda=11,v=1$ and tune $J$ to go from the symmetric phase to the CDW phase ($J_v>0,J_u=0$) or to the spin nematic phase ($J_u>0,J_v=0$). We obtained the phase diagram by calculating Binder ratios and correlation ratios (defined below). The three relevant order parameters are $m_{stripe}\sim \phi_{\eta}$, $m_{O(2)}\sim \frac{\sum_{\eta}(\phi_{\eta})^2}{|\sum_{\eta}(\phi_{\eta})^2|}$ and $m_{O(3)}\sim \frac{Q}{\sqrt{\Tr[Q^2]}}$. From these we define
\begin{eqnarray}
	m_{stripe}^2=\sum_{\eta}|\sum_{\r,\tau}\phi_{\eta}(\r,\tau)|^2 \nonumber\\
	m_{O(2)}^2=\left\lvert\sum_{\r,\tau}\frac{\sum_{\eta}(\phi_{\eta}(\r,\tau))^2}{\sum_{\eta}|\phi_{\eta}(\r,\tau)|^2}\right\rvert^2 \nonumber\\
	m_{O(3)}^2= \Tr[(\sum_{\r,\tau}\frac{Q(\r,\tau)}{\sqrt{\Tr[Q(\r,\tau)^2]}})^2]\,,
\end{eqnarray}
and the corresponding Binder ratios and correlation ratios:
\begin{eqnarray}
	R_B&=&\frac{\left\langle m^4\right\rangle}{\left\langle m^2\right\rangle^2 } \nonumber\\
	R_c&=&\frac{\left\langle \chi(\Q+dq)\right\rangle}{\left\langle \chi(\Q)\right\rangle }
\end{eqnarray}
Here, $\chi(\q)=\sum_{\r} e^{i\q\cdot \r} m(\r) m(0)$ is the Fourier transformation of the order parameter correlation function, $\Q$ is the ordering vector ($\Q=(0,0)$ here), and $dq = (2\pi/L,0)$. 

The phase diagram is shown in Fig.~\ref{bosonicPD} below. The left region corresponds to the spin nematic (SN), and the right region corresponds to the CDW. The blue lines correspond to continuous $O(2)$ transitions, and the orange lines to continuous $O(3)$ transitions. The yellow dashed line most likely corresponds to a first order transition (as indicated by large jumps in the Binder and correlation ratios). The second order behavior becomes more pronounced near the emerged O(4) point where both $J_u$ and $J_v$ are zero (indicated with a star in Fig.~\ref{bosonicPD}). For the continuous $O(2)$ and $O(3)$ transitions we use the correlation ratios obtained at $L=6,8,10,12$ to obtain the transition points indicated in the figure. We obtain very good data collapses using the critical exponents of the $O(2)$ and $O(3)$ universality classes (not shown).
\begin{figure}[htp!]
    \includegraphics[scale=0.5]{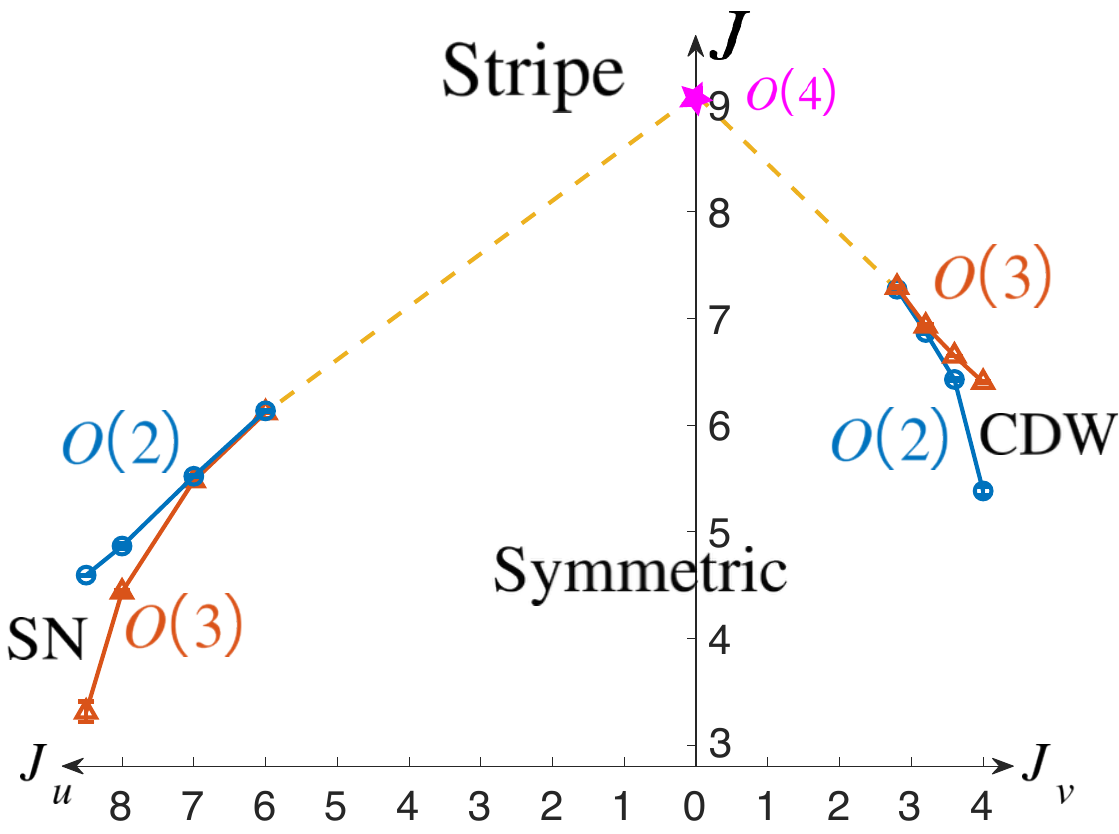}
    \caption{Phase diagram obtained by tuning $J$ and $J_u$ or $J_v$. `Stripe' labels the SDW region, `SN' refers to spin nematic, and `CDW' labels the region with charge density wave order but no spin order. At $J_u=J_v=0$, the pink star labels the $O(4)$ transition point on the $J$-axis between the SDW and the symmetric state. Blue (red) lines indicate continuous O(2) (O(3)) transitions, determined by Binder and correlation ratios. The dashed lines are drawn by hand at locations where both the Binder ratio and the correlation ratio jump, indicating a first order transition.}\label{bosonicPD}
\end{figure}

\section{The $\frac{O(3)\times O(2)}{\mathbb{Z}_2}$ boson-fermion model and the simplified $O(3)$ boson-fermion model}
\subsection{Introduction of the $\frac{O(3)\times O(2)}{\mathbb{Z}_2}$ boson-fermion model}
We now couple the boson model of the previous appendix with fermions by adding following term to the action:
    \begin{eqnarray}
	S_F&=&\int d\tau [\sum_{\r,\r';s,l=\pm}\bar{\psi}_{s,l}(\r,\tau)(\delta_{\r,\r'}\partial_\tau-H(\r,\r')-\delta_{\r,\r'}\mu)\psi_{s,l}(\r',\tau) \nonumber\\
	&+& g\sum_{\r,\eta,s,s',l}(-1)^{r_x+r_y+l}\bar{\psi}_{s,l}(\r,\tau)\sigma_{s,s'}^{\eta}\psi_{s',l}(\r,\tau)\cos(Q_s r_x\theta_\eta)|\phi_{\eta}(\r,\tau)|]	\nonumber\\
	&=&\int d\tau [\sum_{\r,\r';s,l=\pm}\bar{\psi}_{s,l}(\r,\tau)(\delta_{\r,\r'}\partial_\tau-H(\r,\r')-\delta_{\r,\r'}\mu)\psi_{s,l}(\r',\tau) \nonumber\\
	&+& g\sum_{\r,\eta,s,s',l}(-1)^{r_x+r_y+l}\bar{\psi}_{s,l}(\r,\tau)\sigma_{s,s'}^{\eta}\psi_{s',l}(\r,\tau)(\cos(Q_sx)\Re[\phi_{\eta}(\r,\tau)]+\sin(Q_sr_x)\Im[\phi_{\eta}(\r,\tau)])]\,,
\end{eqnarray}
where $l=\pm$ labels two different layers, each containing an identical square lattice, and $s$ labels spin. The Hamiltonian for the fermions is given by
\begin{equation}
	H(\r,\r')=-t(\delta_{\r',\r+dx}+\delta_{\r',\r+dy})-t'(\delta_{\r',\r+dx+dy}+\delta_{\r',\r+dx-dy})+h.c.
\end{equation}
The electron-boson model has a time-reversal symmetry which acts trivially on the bosons, and acts as follows on the fermions:
\begin{eqnarray}
	\psi \rightarrow \gamma^x i\sigma^{y}\psi
\end{eqnarray}
Here, $\gamma^x$ is the Pauli-x matrix acting on the layer index, and $\sigma^y$ acts on the spin indices. 

The full partition function can be written as
\begin{eqnarray}
	Z&=&\int D[\phi^*]D[\phi]e^{-S_B}\Tr_{\{\phi^*,\phi\}}(e^{-\beta H_F}) \\
	&=&\int D[\phi^*]D[\phi]\prod_{\tau=0}^{\beta} e^{-S_B(\tau)\Delta\tau} e^{J\sum (\Re[\phi(\tau+\Delta\tau)]\Re[\phi(\tau)]+\Im[\phi(\tau+\Delta\tau)]\Im[\phi(\tau)])/c^2\Delta\tau}\det(I+\prod_{\tau=0}^{\beta} e^{-\Delta\tau H_F(\tau)}) \nonumber\\
	&=&\int D[\phi^*]D[\phi]\prod_{\tau=0}^{\beta} e^{-S_B(\tau)\Delta\tau} e^{J\sum(\Re[\phi(\tau+\Delta\tau)]\Re[\phi(\tau)]+\Im[\phi(\tau+\Delta\tau)]\Im[\phi(\tau)])/c^2\Delta\tau}|\det(I+\prod_{\tau=0}^{\beta} e^{-\Delta\tau H_F^{l=+}(\tau)})|^2 \nonumber\,,
\end{eqnarray}
where in the last step we have used the anti-unitary time-reversal symmetry.
Together with the real bosonic action, we see that the positive fermion determinant allows for sign problem-free sampling. Another consequence of the Kramers' time-reversal symmetry is that superconductivity must necessarily occur in the $c^{\dagger}\gamma^{x}\sigma^{y} c^{\dagger}$ channel~\cite{Zhang2025}, which corresponds to an inter-layer and spin-singlet local pairing.

\subsection{Quantum Monte Carlo update}
For our simulations we combine the local update together with the Wolff cluster update described in the previous appendix to sample the system efficiently. For the local update of the bosonic field, the ratio is
\begin{eqnarray}
	R_B&=&e^{-S_B'
	(\tau)\Delta\tau+S_B(\tau)\Delta\tau}e^{-J\sum(\Re[\phi(\tau+\Delta\tau)+\phi(\tau-\Delta\tau)]\Re[\phi'(\tau)-\phi(\tau)]+\Im[\phi(\tau+\Delta\tau)+\phi(\tau-\Delta\tau)]\Im[\phi'(\tau)-\phi(\tau)])/c^2\Delta\tau}
\end{eqnarray}
And the contribution for the ratio from fermionic part is
\begin{eqnarray}
	R_F=\left\lvert\frac{\det(I+\prod_{\tau=0}^{\beta} e^{-\Delta\tau H_F'^{l_+}(\tau)})}{\det(I+\prod_{\tau=0}^{\beta} e^{-\Delta\tau H_F^{l_+}(\tau)})}\right\rvert^2
\end{eqnarray}
which can be computed with fast update routine based on the Green's function $G(\tau,\tau)=\langle c(\r,\tau)c^{\dagger}(\r',\tau)\rangle$.
\begin{eqnarray}
	\hat{\Delta}&=&e^{\Delta\tau g\sum_{\r,\eta,s,s',l}(-1)^{r_x+r_y+l}c^{\dagger}_{s,l}(r,\tau)\sigma_{s,s'}^{\eta}c_{s',l}(\r,\tau)(\cos(Q_sr_x)\Re[\phi'_{\eta}(\r,\tau)]+\sin(Q_sr_x)\Im[\phi'_{\eta}(\r,\tau)])}	\nonumber\\
	&\cdot&e^{-\Delta\tau g\sum_{\r,\eta,s,s',l}(-1)^{r_x+r_y+l}c^{\dagger}_{s,l}(\r,\tau)\sigma_{s,s'}^{\eta}c_{s',l}(\r,\tau)(\cos(Q_sr_x)\Re[\phi_{\eta}(\r,\tau)]+\sin(Q_sr_x)\Im[\phi_{\eta}(\r,\tau)])}-I \nonumber\\
	R_F&=&\det(I+\Delta(I-G(\tau,\tau))) \nonumber\\
	G'(\tau,\tau)&=&G(\tau,\tau)(I+\Delta(I-G(\tau,\tau)))^{-1}
\end{eqnarray}
Here $\Delta$ is the matrix form of $\hat{\Delta}$ in single particle basis. The calculation of $R_F$ and $G'(\tau,\tau)$ can be done with a cost that scales as $L^{2}$ because of the sparsity of the $\Delta$ matrix. For the Wolff update with fermions, the probability of flipping the cluster is determined by the ratio of the fermion determinants before and after the cluster flip. 

\subsection{Introduction of the simplified $O(3)$ boson-fermion model}
When simulating the above model with small $Q_s$ (we have tried $Q_s = 1/8$), we find that with sizable coupling strengths, the feedback from the fermions on the bosons cannot be neglected, and changes the phase diagram in important ways. In particular, we find that when $g$ is order one, the bosons tend to order at momentum $-Q_s$, even though the gradients in $S_B$ would like them to order at zero momentum, and phase transitions become strongly first order. To understand this, note that integrating out the fermions will introduce following symmetry-allowed term for the bosons:
\begin{equation}\label{gammaterm}
- \gamma \sum_\r \sum_\eta \left[\Re(e^{iQ_sr_x} \phi_\eta(\r))\right]^2\,,
\end{equation}
where on physical grounds we expect $\gamma>0$, as this favors the $\phi$ field to order at momentum $-Q_s$, which makes that at mean-field level the electron-boson coupling term looks like a conventional anti-ferromagnetic order parameter for the fermions. In principle one can weaken or even completely remove this effect by adding an explicit counter term to the action which is of the same form as in Eq.~\eqref{gammaterm}, but with opposite sign. However, in practice it is difficult to know the correct magnitude one should take for this counter term.

Instead of adding the counter term, we choose to adopt a simpler way around the finite-momentum ordering of the bosons. We put $Q_s=0$ and replaced the O(2) symmetry by $\mathbb{Z}_2$, such that the symmetry of the electron-boson model becomes (O(3)$\times\mathbb{Z}_2)/\mathbb{Z}_2 =$ O(3). With this modification, the $J_v$ term drops out of the action, but we can keep the $J_u$ term. Our model thus becomes equivalent to $2+1$D classical O(3) model coupled to fermions, but with an additional (and important) $J_u$ term promoting spin nematicity. This modification is easily implemented in practice by putting the imaginary part of $\phi$, and also $Q_s$, to zero in the electron-boson model described above. 

\begin{figure*}[htp!]
  \includegraphics[scale=0.5]{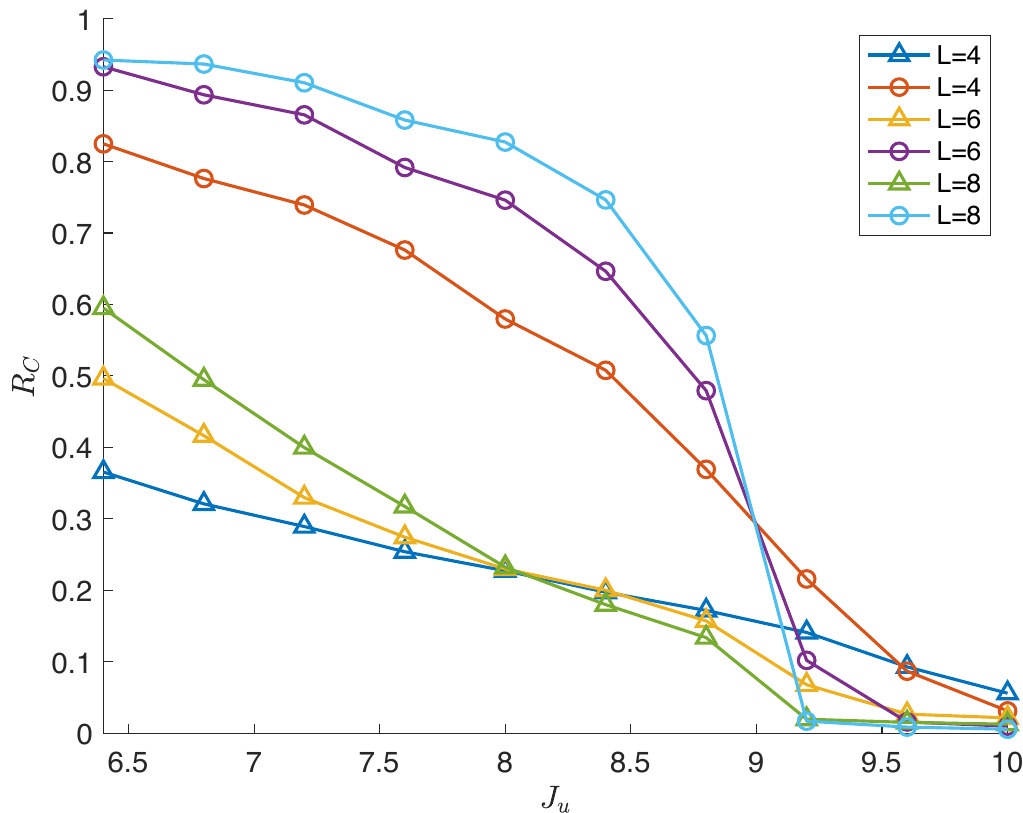}
  \caption{Phase diagram identified from correlation ratio $R_C$ crossing for $O(3)$ order (lines with triangles) and AFM order (lines with circles). From left to right separated with two crossing points, the phases are disorder, SN and AFM.}\label{O3PD}
\end{figure*}

Because we take $Q_s=0$, the SDW becomes a conventional anti-ferromagnet (AFM). The transition from the spin nematic to the AFM is found to be a continuous $\mathbb{Z}_2$ or Ising transition. Concretely, we have used the parameters $(t,t',\mu,J_u,g,\beta)=(2,-0.5,1.6,2J+4,\sqrt{J},2L)$, and found the phase diagram as a function of $J$ shown in Fig.~\ref{O3PD}. This figure shows the correlation ratios for both the AFM and spin nematic order parameters obtained at different system sizes. From these curves we can indeed identify two separate phase transition where first spin nematic order develops (broken $O(3)/\mathbb{Z}_2 = RP^2$ symmetry), and then AFM order (broken $O(3)$ symmetry). We see that the positions of the phase transition points are similar to those of the purely bosonic model discussed above. The Fermi surface reconstruction happens at the transition from spin nematic to AFM. If we use larger electron-boson coupling strengths, we find the SN-AFM transition to merge with the $O(3)$ transition and form a single first-order transition between the AFM and the symmetric phase.

\subsection{Fermi surface in the spin nematic phase}
\begin{figure*}[htp!]
	\includegraphics[scale=0.5]{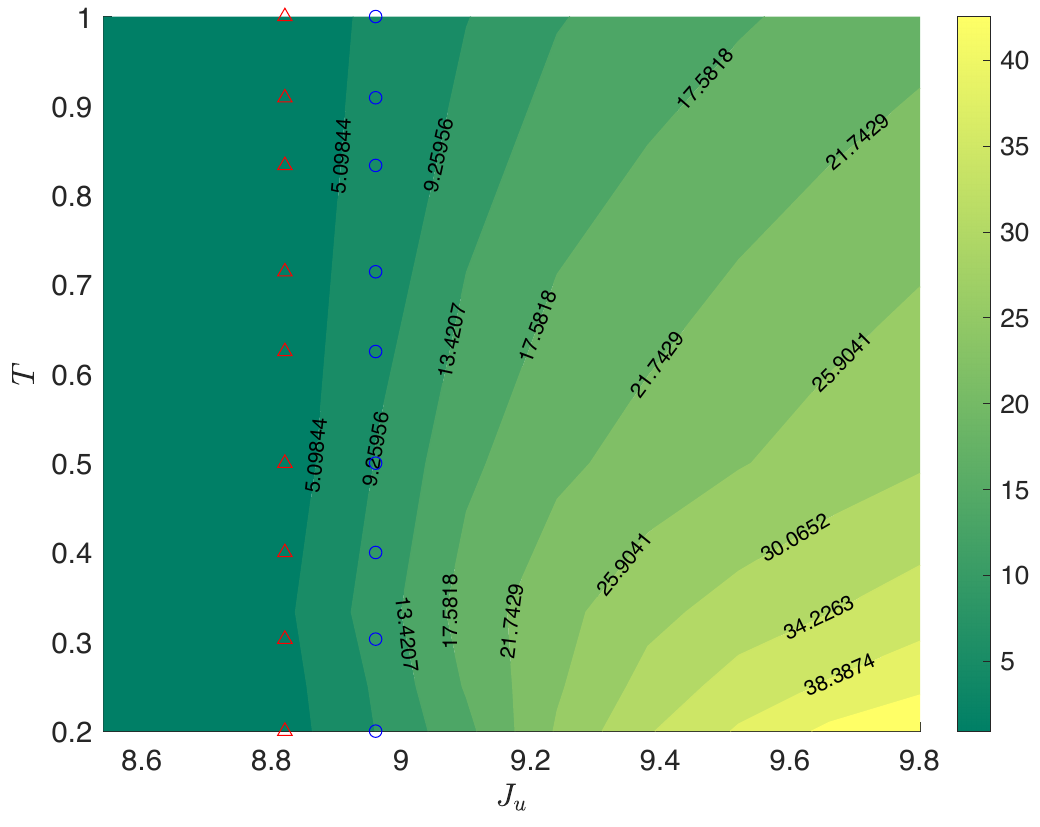}
	\caption{The AFM correlation length at finite temperature for $L=12$. The left region is SN phase and the right region corresponds to the AFM. The red triangles and blue circles label points at respectively $[J,J_u]=[2.41,8.82]$ and $[J,J_u]=[2.48,8.96]$ where the spectral weight shown in Figs.~\ref{FS1} and~\ref{FS2} was calculated (spectral weight calculations were done with $L=16$).}
    \label{AFMlength}
\end{figure*}

\begin{figure*}[htp!]
	\includegraphics[scale=0.3]{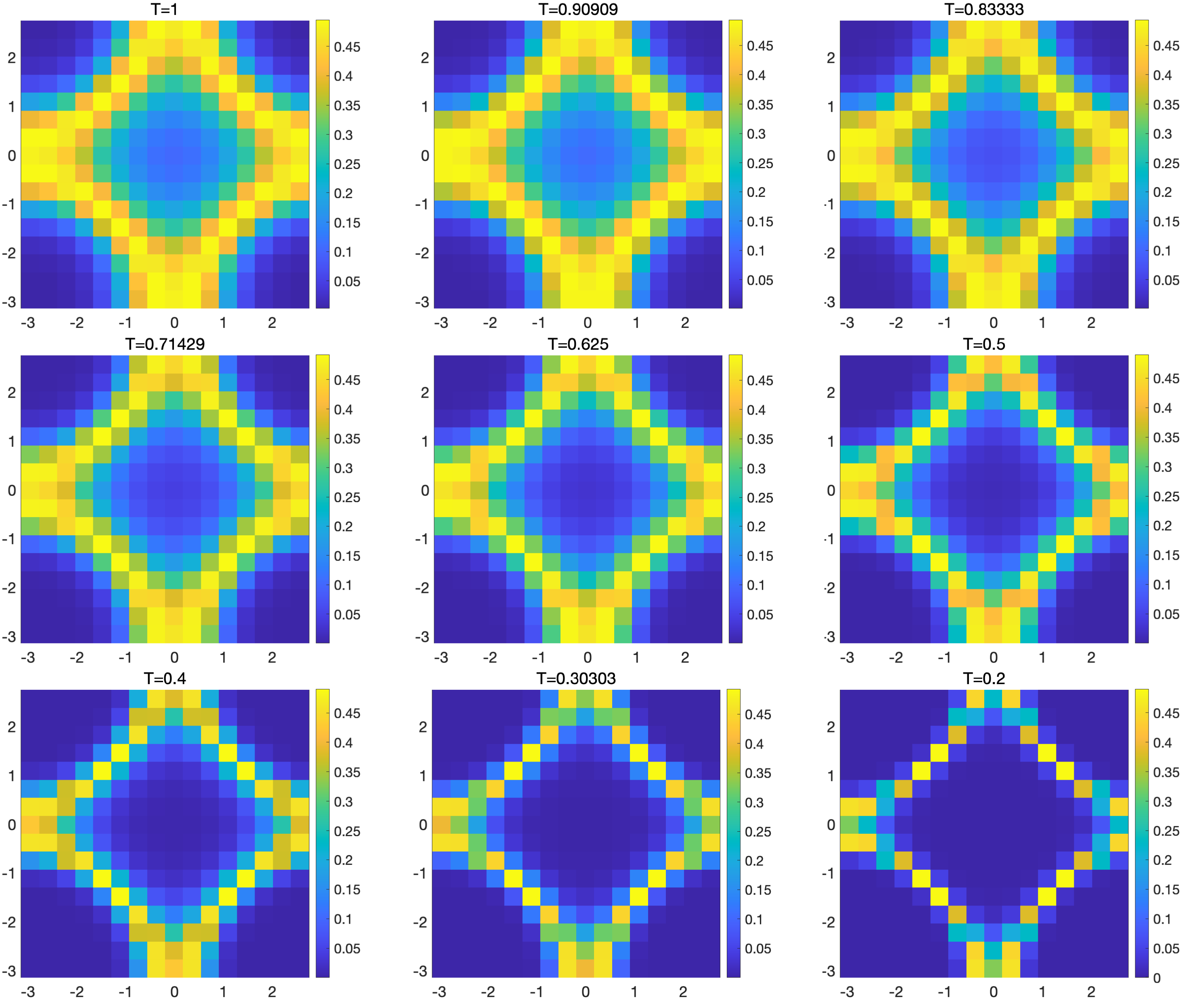}
	\caption{$G(\k,\tau=\beta/2)$ as a proxy for the finite-temperature spectral weight around the Fermi surface, obtained with $L=16$ and $J_u=8.82$.}\label{FS1}
\end{figure*}

\begin{figure*}[htp!]
	\includegraphics[scale=0.3]{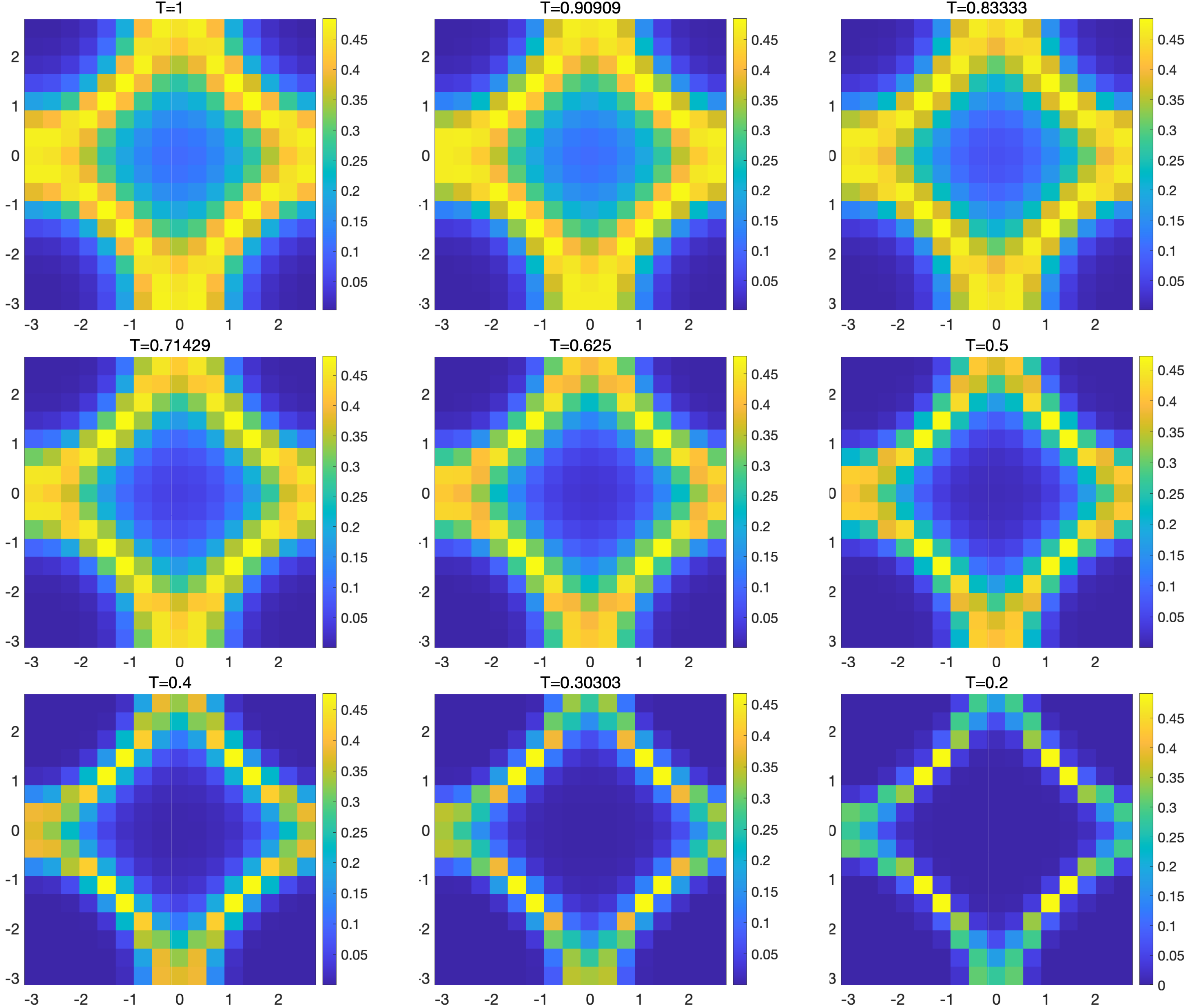}
	\caption{$G(\k,\tau=\beta/2)$ as a proxy for the finite-temperature spectral weight around the Fermi surface, obtained with $L=16$ and $J_u=8.96$. }\label{FS2}
\end{figure*}

Next we zoom in on the parameter region from $[J,J_u]=[2.27,8.54]$ to $[J,J_u]=[2.9,9.8]$. In Fig.~\ref{AFMlength} we plot the AFM correlation length at finite temperature for $L=12$. This correlation length was obtained via the relation $\xi_{AFM}=\frac{L}{2\pi}\sqrt{\frac{1}{R_C}-1}$~\cite{Carsten2020,Chaikin_Lubensky_1995}. Finally, we have also calculated the imaginary-time fermion propagator $G(\k,\tau=\beta/2)$, which is a proxy for the fermion spectral weight coming from states in a window of order $T$ around the Fermi energy~\cite{Trivedi1995,Schattner2016}. We calculated $G(\k,\tau=\beta/2)$ at two different parameter values $[J,J_u]=[2.41,8.82]$ and $[J,J_u]=[2.48,8.96]$, for different temperatures and with system size $L=16$. The points in the $(T,J_u)$ plane at which $G(\k,\tau=\beta/2)$ was calculated are indicated with red triangles and blue circles in Fig.~\ref{AFMlength}. From Figs.~\ref{FS1} and~\ref{FS2} we clearly recognize a large Fermi surface in the spin nematic phase, with the spectral weight in the nodal regions (i.e. near $(\pi,0)$ and $(0,\pi)$) gradually becoming weaker as the temperature is lowered and the transition to the AFM is approached.


\section{The plaquette-Ising model}
In this final appendix we provide some additional details on the plaquette-Ising model of the main text. Let us first write out its Hamiltonian more explicitly as
\begin{eqnarray}
    H&=&H_f+H_I \\
    &=&-t\sum_{\r}(c^{\dagger}_{\r}c_{\r+dx}+c^{\dagger}_{r}c_{\r+dy}+h.c.)-t'\sum_{\r}(c^{\dagger}_{r}c_{\r+dx+dy}+c^{\dagger}_{r}c_{\r+dx-dy}+h.c.)+\mu\sum_{\r}c^{\dagger}_{\r}c_{\r} \nonumber\\
    &-&g \sum_{\R} (c^{\dagger}_{\R} \sigma^z c_{\R}-c^{\dagger}_{\R+dx} \sigma^z c_{\R+dx}+c^{\dagger}_{\R+dx+dy} \sigma^z c_{\R+dx+dy}-c^{\dagger}_{\R+dy} \sigma^z c_{\R+dy}) S_\R -J \sum_{\R}(S_{\R}S_{\R+2dx}+S_{\R}S_{\R+2dy}), \nonumber
\end{eqnarray}
where $S_{\R}=\pm1$ are classical Ising spins living in on one quarter of the plaquettes of the square lattice. These Ising spins couples to the z-component of the electron spins at the four corners of the plaquettes with alternating sign. As in the main text we have used the parameters $(t,t',\mu,J)=(1,-0.35,1.4,0.1)$ in our simulations. 

\begin{figure*}[htp!]
  \includegraphics[scale=0.5]{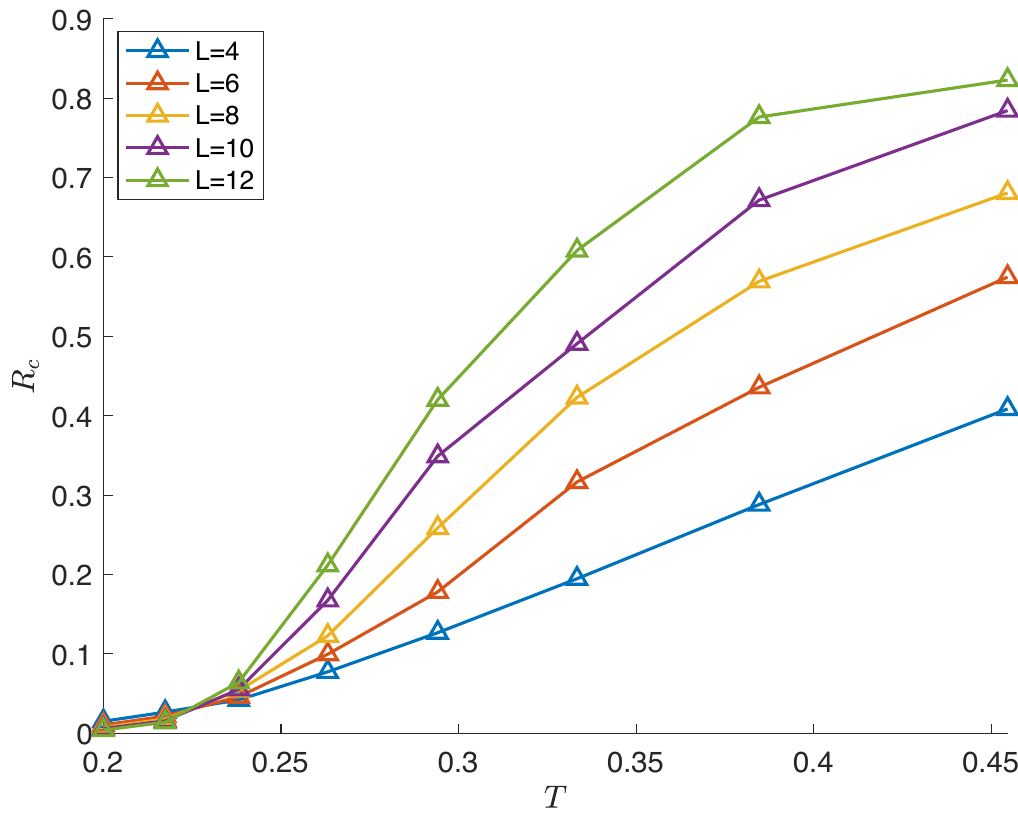}
  \caption{Benchmark result for $g=0$ case, where a cross between $T\in(0.22,0.23)$ indicates the 2D Ising transition.}\label{decoupled}
\end{figure*}

As a first benchmark we have checked the decoupled $g=0$ case, which is the pure ferromagnetic Ising model on a 2D square lattice. According to Onsager's exact solution, the phase transition happens at a critical temperature $T_c=\frac{2J}{\ln(1+\sqrt{2})}\approx0.227$. We have used the correlation ratio defined as $R_c=\frac{\langle\chi(\Q+dq)\rangle}{\langle\chi(Q)\rangle}$ to numerically find this phase transition point. Here $\chi(\Q)=\sum_{\r} e^{i\Q\cdot \r} \langle m(\r) m(0) \rangle$ is the Fourier transformation of the order parameter correlation function, with $m(\r)=c^\dagger_\r\sigma^z c_\r$, and $dq=(2\pi/L,0)$. The result is shown in the Fig.~\ref{decoupled}. We see that the $R_c$ curves obtained at different $L$ cross at $T\sim 0.227$, as expected.

\begin{figure*}[htp!]
  \includegraphics[scale=0.5]{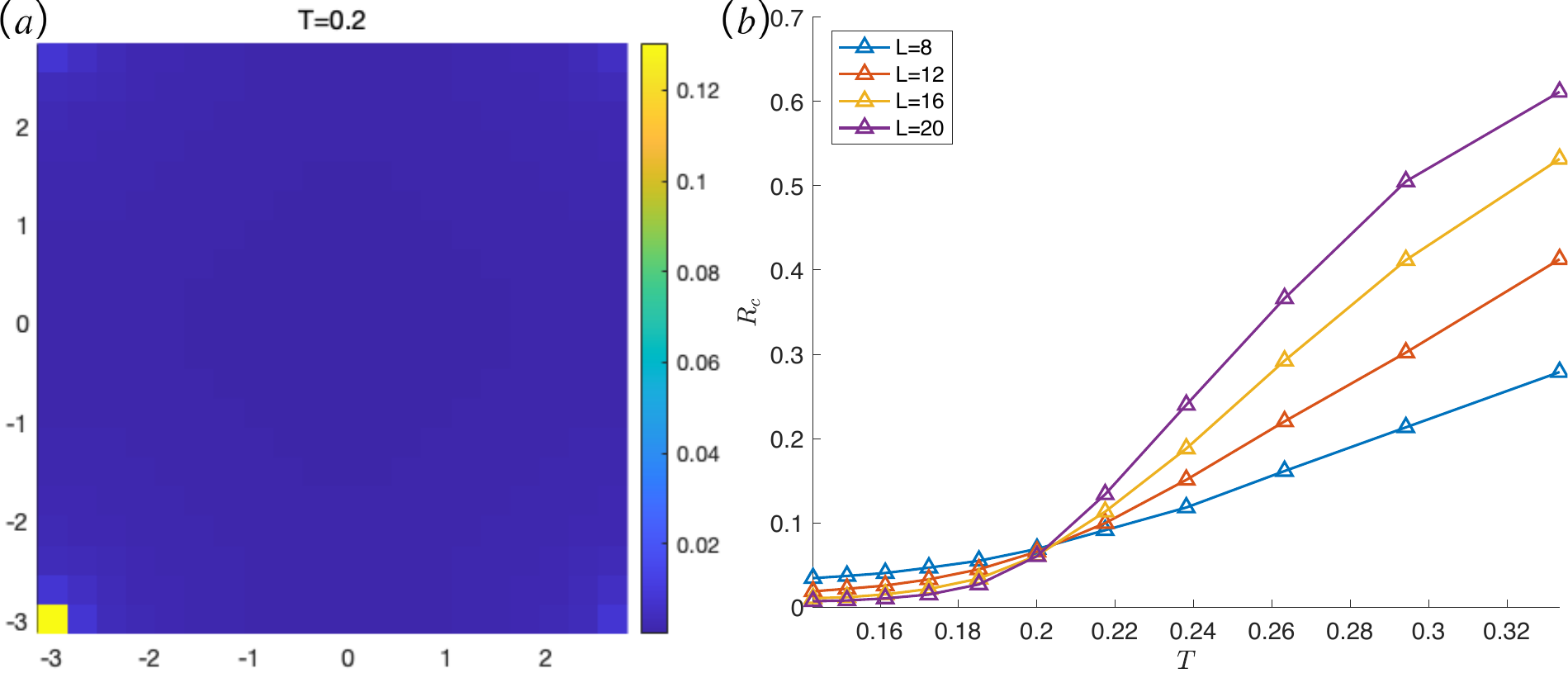}
  \caption{Results for the plaquette-Ising model obtained with $g=1$. (a) $\chi(\q)$ showing a peak at the AFM wavevector $\Q=(\pi,\pi)$. (b) Correlation ratio curves obtained at different $L$. The crossing indicates a critical temperature $T_c\sim 0.2$.}\label{g1}
\end{figure*}

Next we turn on the interaction between the Ising spins and the fermions. In Fig.~\ref{g1} we plot $R_c$ curves obtained with $g=1$, which show a crossing around a temperature $T\sim 0.2$. We thus find that by taking $g=1$ the phase transition shifts to smaller temperatures, in contrast to most cases where the transition temperature actually increases by coupling to fermions~\cite{Schattner2016,Xu2017,Liu2018,Liu2019}. In Fig.~\ref{g1} we also show $\chi(\q)$, which has a clear peak at $\Q = (\pi,\pi)$, confirming the AFM ordering of the fermions.

Let us now try to understand the shift of $T_c$ to lower temperatures when $g>0$.  As a warm-up, first consider the conventional on-site spin-fermion coupling:
\begin{equation}
    H_I'=-g \sum_{\r} (-1)^{r_x+r_y}(c^{\dagger}_{\r} \sigma^z c_{\r}) S_\r \label{onsite}
\end{equation}
Note that here the Ising spins live on the same lattice as the electrons. The RKKY interaction comes from the linear response of the electron spins to the coupling with the Ising spins. We can write the effective RKKY interaction as
\begin{equation}
    H_{RKKY}'=\frac{1}{2}\sum_{\r,\r'} J'_{RKKY}(\r-\r') S_\r S_{\r'}\,, 
\end{equation}
with
\begin{eqnarray}
    J'_{RKKY}(\r-\r')=-J(\r)J(\r')\chi(\r-\r')\,,
\end{eqnarray}
where $J(\r)=-g(-1)^{r_x+r_y}$ is the local spin-fermion coupling, and $\chi(\r-\r')$ is the free electron magnetic susceptibility. The latter is easily obained by using Wick's theorem as
\begin{eqnarray}
    \chi(\r-\r')&=&\langle c^{\dagger}_{\r}\sigma^{z} c_{\r}c^{\dagger}_{\r'}\sigma^zc_{\r'}\rangle \\
    &=&-\langle c_{\r,\uparrow}c^{\dagger}_{\r',\uparrow}\rangle\langle c_{\r',\uparrow}c^{\dagger}_{\r,\uparrow}\rangle-\langle c_{\r,\downarrow}c^{\dagger}_{\r',\downarrow}\rangle\langle c_{\r',\downarrow}c^{\dagger}_{\r,\downarrow}\rangle \nonumber\\
    &=&-|G_{\uparrow}(\r,\r')|^2-|G_{\downarrow}(\r,\r')|^2 \nonumber
\end{eqnarray}
Here we assume $\r\neq \r'$, and defined the equal-time fermion Green's function $G_{\alpha}(\r,\r')=\langle c_{\r,\alpha}c^{\dagger}_{\r',\alpha}\rangle$. In the final expression we have used the $\mathbb{Z}_2$ spin flip symmetry and hermiticity of $G_{\alpha}(\r,\r')$. We thus find that
\begin{equation}
    J'_{RKKY}(\r-\r')=\frac{1}{2}g^2\sum_{\r,\r'} (-1)^{(r_x-r'_x)+(r_y-r'_y)}(|G_{\uparrow}(\r,\r')|^2+|G_{\downarrow}(\r,\r')|^2)S_\r S_{\r'}.
\end{equation}
This is an interaction with alternating sign, which is negative for nearest neighbors. Since the amplitude of the Green's function decays algebraically, the nearest neighbor term dominants, promoting ferromagnetism for the Ising spins and hence shifting the phase transition to \emph{higher} temperatures compared to the decoupled ($g=0)$ Ising model. 

\begin{figure*}[htp!]
  \includegraphics[scale=0.5]{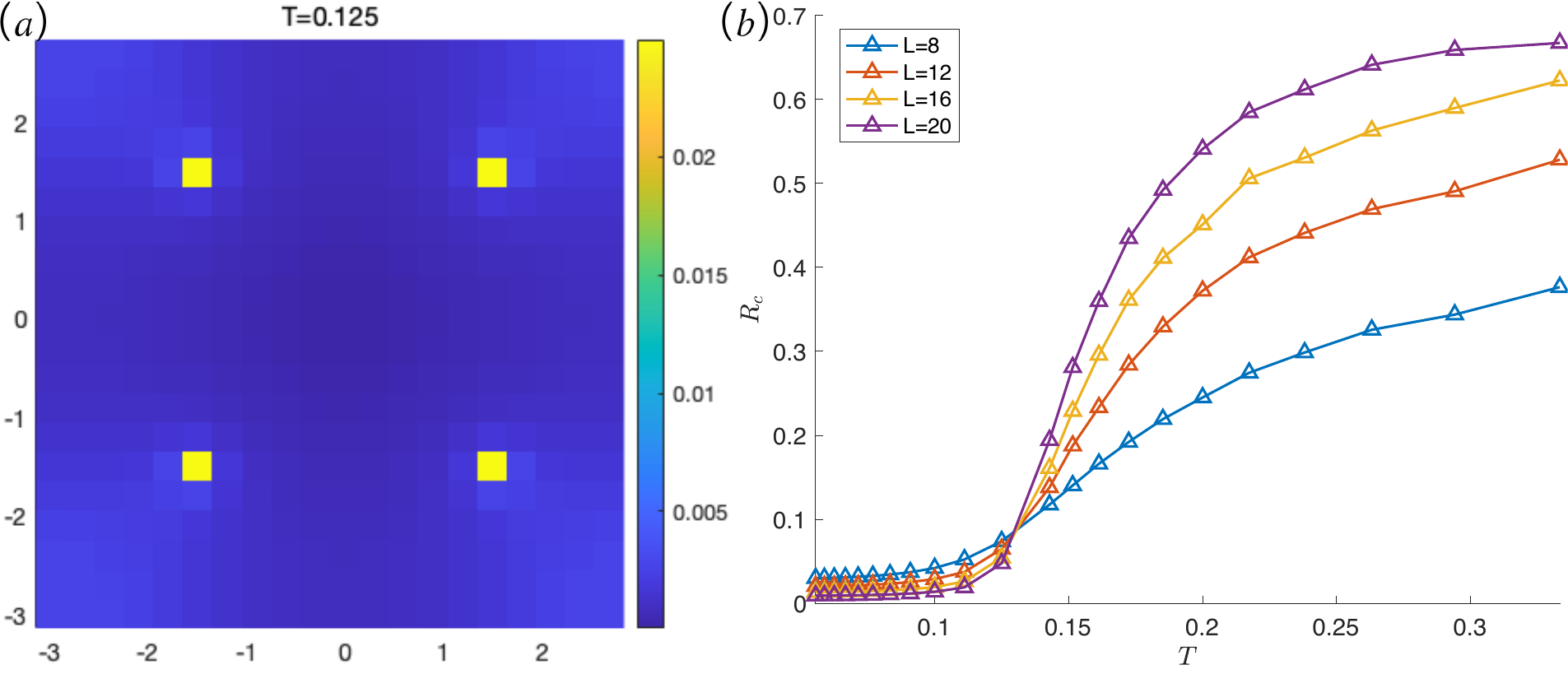}
  \caption{Results for the plaquette-Ising model obtained with $g=1$ and $J=0$.  (a) $\chi(\q)$ showing a peak at $\q=(\pm\pi/2,\pm\pi/2)$. (b) Correlation ratio curves obtained at different $L$. The crossing indicates a critical temperature $T_c\sim 0.13$.}\label{Jzero}
\end{figure*}

Let us now come back to the plaquette-Ising model, which crucially has a different spin-fermion coupling than then the purely on-site coupling of Eq.~\eqref{onsite}. Written out in detail, it is given by
\begin{eqnarray}
    H_I=-g \sum_{\R} (c^{\dagger}_{\R} \sigma^z c_{\R}-c^{\dagger}_{\R+dx} \sigma^z c_{\R+dx}+c^{\dagger}_{\R+dx+dy} \sigma^z c_{\R+dx+dy}-c^{\dagger}_{\R+dy} \sigma^z c_{\R+dy}) S_\R
\end{eqnarray}
So the Ising spin couples to the `plaquette moment' $m(\R)=c^{\dagger}_{\R} \sigma^z c_{\R}-c^{\dagger}_{\R+dx} \sigma^z c_{\R+dx}+c^{\dagger}_{\R+dx+dy} \sigma^z c_{\R+dx+dy}-c^{\dagger}_{\R+dy} \sigma^z c_{\R+dy}$, which has a free fermion susceptibility
\begin{eqnarray}
    \chi(\R-\R')=&&\langle m(\R) m(\R')\rangle \\
    =&-&\sum_{\alpha=\uparrow,\downarrow}|G_{\alpha}(\R,\R')|^2+|G_{\alpha}(\R+dx,\R'+dx)|^2 \nonumber\\
    &+&|G_{\alpha}(\R+dx+dy,\R'+dx+dy)|^2+|G_{\alpha}(\R+dy,\R'+dy)|^2 \nonumber\\
    &-&|G_{\alpha}(\R,\R'+dx)|^2-|G_{\alpha}(\R+dx,\R')|^2 \nonumber\\
    &+&|G_{\alpha}(\R,\R'+dx+dy)|^2+|G_{\alpha}(\R+dx+dy,\R')|^2 \nonumber\\
    &-&|G_{\alpha}(\R,\R'+dy)|^2-|G_{\alpha}(\R+dy,\R')|^2 \nonumber\\
    &-&|G_{\alpha}(\R+dx,\R'+dx+dy)|^2-|G_{\alpha}(\R+dx+dy,\R'+dx)|^2 \nonumber\\
    &+&|G_{\alpha}(\R+dx,\R'+dy)|^2+|G_{\alpha}(\R+dy,\R'+dx)|^2 \nonumber\\
    &-&|G_{\alpha}(\R+dx+dy,\R'+dy)|^2-|G_{\alpha}(\R+dy,\R'+dx+dy)|^2.
\end{eqnarray}
Plugging this expression into the RKKY interaction term $H_{RKKY}=-\frac{1}{2}g^2\sum_{\R,\R'} \chi(\R-\R')S_\R S_{\R'}$ we see that the interaction between nearest neighbor spins receives both positive and negative contributions. In particular, for a pair of neighboring Ising spins, the RKKY interaction receives 2 FM contributions from electrons removed by distance 1, 2 AFM contributions from electrons at distance $\sqrt{2}$, 4 AFM contributions from electrons at distance $2$, 4 FM contributions at distance $\sqrt{5}$, 2 FM contributions at distance $3$, and finally 2 AFM contributions at distance $\sqrt{10}$. It is now much harder to predict what will be the net effect of all these different contributions. We have therefore simulated the plaquette-Ising model with $J=0$, to see what is the ordering tendency preferred by the RKKY interaction. The results obtained with $J=0$ are shown in Fig.~\ref{Jzero}. From the correlation ratios we find a magnetic transition at $T\sim 0.13$. We also see that $\chi(\q)$ develops clear peaks at $\q=(\pm\pi/2,\pm\pi/2)$. As this is the spin structure factor for the electrons, these momenta indicate \emph{anti-ferromagnetic} ordering of the Ising spins. This shows that the RKKY interaction counteracts the ferromagnetic Ising coupling $J$ in the Hamiltonian, hence lowering $T_c$.

\end{document}